\documentclass[aps,prb,superscriptaddress,twocolumn]{revtex4-2}
\usepackage{amsmath}
\usepackage{amssymb}
\usepackage{amsfonts}
\usepackage{color}
\usepackage{graphicx}
\usepackage{amsfonts}
\usepackage{textcomp}
\newcommand{\beq}{\begin{equation}}
\newcommand{\eeq}{\end{equation}}
\newcommand{\bea}{\begin{eqnarray}}
\newcommand{\eea}{\end{eqnarray}}

\newcommand{\cb}{\color{blue}}

\begin{document}
\title{Phase separation in two-dimensional electron systems: Experimental view}

\author{V.~M.~Pudalov}
\affiliation{
V.~L.~Ginzburg Research Center at P.~N.~Lebedev Physical Institute,  Moscow 119991, Russia.\\
HSE University, Moscow 101000, Russia
}

\setcounter{section}{0}

\begin{abstract}
Key experimental results on unveiling and studying properties of a multiphase state that arises in two-dimensional electron systems due to the interplay of interelectron interactions and disorder are reviewed. We focus on the experimental results obtained with high  mobility Si- field effects structures (Si-MOS), in which the interaction effects at low carrier concentrations are most pronounced due to the strong e-e interactions,  multi-valley spectrum, and the short-range character of the random potential. The reviewed effects of phase separation include features in transport, magnetotransport and thermodynamics. Consideration of a number of experimental results is supplemented with a brief review of their theoretical interpretation.
\end{abstract}

\maketitle

\tableofcontents

\section{Introduction}
Recently, phase separation effects came to the forestage as the objects of intent 
attention in condensed matter physics.
Hystorically first and the most remarkable phase separation effects were found in
manganites \cite{rama_PRL_2004}, then the phase separation appeared to play essential role in high-Tc 
superconductors \cite{gorbatsevich_1994, kresin-book}, low-dimensional organic crystals \cite{lebed-book}, etc.
To date it became clear that phase separation is ubiquitous  rather than exotics. 

Several comprehensive reviews were published in recent years \cite{fradkin_Annu.RevCondMatPhys_2010, kagan_PhysRep_2021}, they considered mainly theoretical aspects of the physics of phase separation. The current mini-review partially compensates this shortcoming and considers several representative manifestions of the phase separation in  experiments with two-dimensional (2D) systems of interacting electrons.
In all considered examples the driving force behind the phase separation is the competition between disorder and interparticle interactions. It is known that the effects of interaction are the stronger, the lower the dimension of the system.

\section{Two-dimensional electron  systems with strong interactions. Theory overview}
Correlation plays a crucial role for 
electrons with a $1/r$ pair potential moving in a neutralizing
charge background \cite{giuliani_book}. Its importance grows both with lowering
the density and the space dimensionality, and tends to qualitatively change the predictions
of simple schemes, such as the Hartree-Fock HF or random-phase approximation RPA \cite{giuliani_book}. The interaction strength is commonly characterized by the dimensionless parameter $r_s$ - the ratio of the potential interaction energy $E_{ee}$ and kinetic Fermi  energies $E_F$; for electrons in (001)-Si MOS structure  $r_s= 2.63\times (10^{12}/n[$cm$^{-2}])^{1/2}$ \cite{ando_review}. 

In the low-density  
strongly correlated  electron liquid, the energy balance determining the system
properties is played on a very minute scale and, to get meaningful
predictions, a great accuracy such as the one afforded
by quantum Monte Carlo (QMC) methods is necessary \cite{giuliani_book}. In two dimensional systems, the interplay of disorder and electron-electron interactions gives birth to many exciting effects, some of them are considered below. We begin with the  negative compressibility of the electron liquid - the effect  that paves the way for phase separation.

    \subsection{Negative compressibility and phase separation. Analytical results}
 The inverse compressibility (or $\partial \mu/\partial n$) of a system reflects how its electrochemical potential changes with carrier density
    \begin{equation}
    \kappa^{-1}=n^2\frac{\partial^2E_{tot}}{\partial n^2}=n^2\left( \frac{\partial \mu}{\partial n} \right),
    \label{Eq1}
   \end{equation}
    with $n$ being the carrier density, and $\mu$ the  electrochemical potential. 
    For noninteracting electrons $\kappa$ is proportional to the single-particle density of states $D$, which in 2DE systems  is density independent, being $D_2=g_v m/(\pi \hbar^2)$, where $g_v$ - is the valley degeneracy ($g_v=2$ for (001)-Si MOS).
       
    This picture, however, changes drastically
    when interactions are included. It was realized already in the 1980s that compressibility of the 2DES can become negative at low densities, owing to electron-electron interactions \cite{bello_JETP_1981}. Exchange and correlation
    effects weaken the repulsion between electrons, thereby
    reducing the energy cost, thus leading to negative and singular corrections to $\partial \mu/\partial n$.
    At zero magnetic field this effect is due primarily to the exchange energy while at high field  the correlation energy plays a significant role as well \cite{efros_SSC_1988}. 

    Within the Hartree-Fock (HF) theory, which includes both the density-of-states and exchange terms, for a clean system with no disorder
    one gets:
   \begin{equation}
    	\frac{\partial \mu}{\partial n} = \frac{\pi \hbar^2}{m}-\left(\frac{2}{\pi}\right)^{1/2}\frac{e^2}{4\pi\epsilon}\frac{1}{n^{1/2}}
    \end{equation}

Thus, upon decreasing density, the compressibility for a clean system gets negative and  tends to $-\infty$. 
  The sign change of compressibility means that when the concentration of electrons in the system varies, the changes in the potential energy  due to the interelectron interaction, having the opposite sign, exceed the changes in the  kinetic energy.
    Experimentally, the sign change of $(\partial\mu/\partial n)^{-1})$ has been found  first in the capacitance and chemical potential measurements in magnetic field \cite{krav_PLA_1989, krav_PLA_1990, krav_PRB_1990} and later was confirmed in the  measurements performed by the field penetration technique in zero field \cite{eisenstein_1992, eisenstein_1994, shapira_1996} (Sec.~\ref{Field penetration}).

    \subsection{Taking disorder into account }
    
  Disorder is inevitably present in real two-dimensional systems \cite{suris_JETP_1978}.
    No matter how small the potential fluctuations in the most advanced 2D structures are, they lead to a significant change in the behavior of thermodynamics and transport in the system as carrier density decreases.
    
    	\subsubsection{Results in the framework of quantum interaction corrections and renormalization group theory}
    
    Explicit calculations on the ``metallic''  side (high conductivity $\sigma\times(h/e^2)=E_F\tau/\hbar \gg  1$), within the renorm-group  and interaction correction theory,  show no
    singular correction to the compressibility in leading order
    in disorder \cite{fink_JETP_1983, fink_ZPhys_1984, castellani_PRB_1984-1, castellani_PRB_1984-2, altshuler-aronov_review}, and even to the second order \cite{belitz-kirkpatrick_1994}.
    
    There were several     pioneering    efforts \cite{si-varma_1998, pastor-dobro_1999} in
    addressing the interplay between interaction and disorder and their effect in thermodynamic properties, by calculating corrections  to the compressibility from the     exchange and correlation contribution to the ground state energy \cite{si-varma_1998}.
   However, the  predicted   vanishing of the compressibility as the transition to the localized state is approached     from the metallic side 
   is at odd with experimental results (Section \ref{Experimental studies}).
    
  \subsubsection{Numerical results}
  
\begin{figure}
\includegraphics[width=200pt]{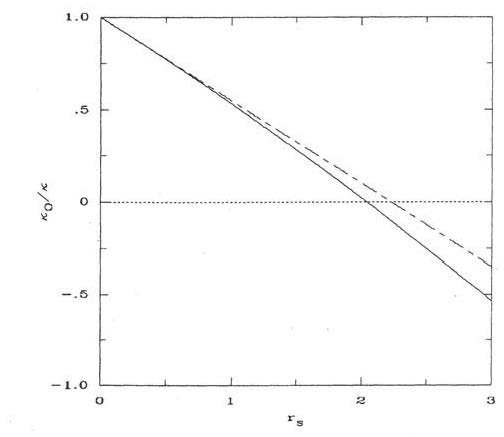}
\includegraphics[width=190pt]{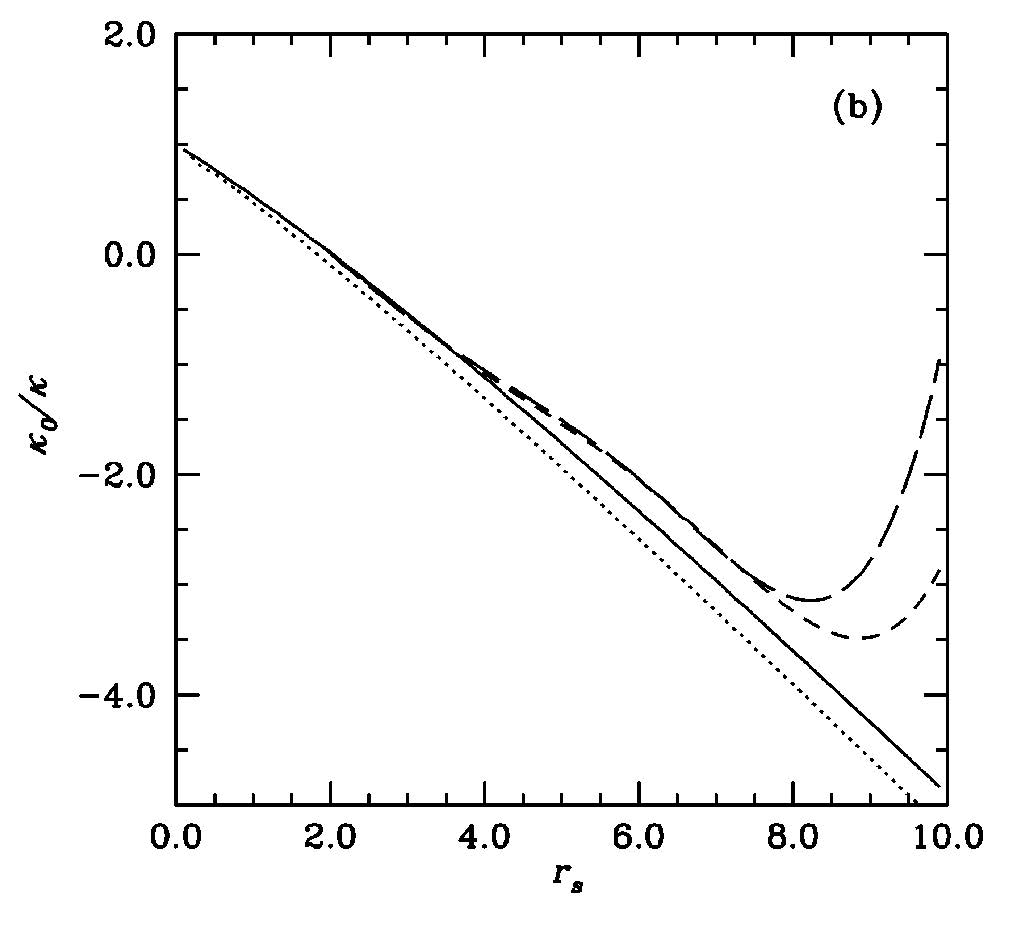}
  	\caption{(a) Inverse compressibility $\kappa_0/\kappa$ of the electron gas as function of density parameter $r_s$   calculated using Eq.~(\ref{tanatar-ceperley-eq17}). 
    Dashed line is the compressibility in the Hartree-Fock approximation. Adapted from Ref.~\cite{tanatar-ceperley_1989}. (b) $\kappa_0/\kappa$ calculated for a wider range of $r_s$. The short and   		long-dashed curves are for impurity densities $n_i=5\times 10^{10}$ and $10^{11}$cm$^{-2}$, respectively. Adapted from Ref.~\cite{asgari-tanatar_2002}.)
  	}
  	\label{compressibility-tanatar&ceperley-asgari&tanatar}
  \end{figure}

  The unlimited  divergency of $\kappa^{-1}$ for the clean 2DE system is cut-off in the presence of disorder. 
  Having obtained the equation of state [i.e., $E(r_s)$] of the normal liquid, Tanatar and Ceperley 
  \cite{tanatar-ceperley_1989} calculated the compressibility using
  \beq
  \frac{\kappa_0}{\kappa} = 1-\frac{\sqrt{2} r_s}{\pi} +\frac{r_s^4}{8} \left[\frac{d^2}{dr^2_s}
  -\frac{1}{r_s}\frac{d}{dr_s}  \right]E_c.
  \label{tanatar-ceperley-eq17}
  \eeq
    Here $\kappa_0=\pi r^4_s /2$ is the compressibility of a noninteracting system (see Fig.~\ref{compressibility-tanatar&ceperley-asgari&tanatar}a), and $E_c$ - correlation energy 
    \cite{tanatar-ceperley_1989} 
 In the above equation, the compressibility becomes negative
  around $r_s=2.03$, slightly before the Hartree-Fock result of 2.22. 
  
  Within the self-consistent HF approximation and the deformable jellium model, Orozco et al. \cite{oronzco_2003} calculated the ground-state compressibility of the 2DE system for the density region which includes the liquid to localized phase transition. 
    A sudden   change in the behavior of the chemical potential and a large divergence in the inverse compressibility have been found for low densities. The change of sign of the inverse compressibility  and its overall behavior versus $r_s$ calculated in this work, agree qualitatively with the behavior of $\kappa^{-1}$ observed in experiments (see Section \ref{Experimental studies}).

  The thermodynamic compressibility can be readily calculated from  the ground-state energy
    \begin{equation}
  	\frac{\kappa_0}{\kappa} = -\frac{r_s^3}{8}\left[\frac{\partial Eg}{\partial r_s}-r_s\frac{\partial^2 E_g}{\partial r_s^2}  \right].
  \end{equation}

Asgari and Tanatar \cite{asgari-tanatar_2002} calculated the ground state energy and compressibility within DFT and dynamical mean field formulation.     The results are presented in  
  Fig.~\ref{compressibility-tanatar&ceperley-asgari&tanatar}b.    The solid curve 
  here shows $\kappa^{-1}$ for a clean system. 
  They considered the disorder effect within two models: (i) a density independent scattering rate  $\gamma$ - similar to Si and Varma  \cite{si-varma_1998}, and (ii) in the mode coupling approximation, with $\gamma$ dependent on $r_s$ through the screened impurity scattering potential.   The dotted curve was   calculated with a constant $\gamma$. It stays negative at low density,
  qualitatively similar to that for the clean system. 
  Most important, $\kappa_0/\kappa$ calculated within the mode-coupling  approximation, which includes the screened   electron-impurity scattering potential, 
  exhibits a minimum and starts rising towards positive values.
  Thus, the inverse compressibility upturn at low densities  is  the effect of disorder solely.  On the other hand, the disorder does not affect $\kappa$  in the range of $r_s =2 \div 4$.
  
The  density at which the calculated inverse compressibility experiences a minimum, depends on the impurity density $n_i$ (see Fig.~\ref{compressibility-tanatar&ceperley-asgari&tanatar}b). In 
 experiments \cite{ilani_PRL_2000, ilani_2001, dultz&jiang_PRL_2000}  the inverse compressibility also shows an upturn after going through a minimum. Initially \cite{dultz&jiang_PRL_2000}, this minimum has been suggested as a thermodynamic signature of the metal-insulator transition.  However, later on the two effects were disentangled and the anomalous behavior
 of $1/\kappa$  was attributed to the inhomogeneous nature of the insulating phase, as  demonstrated experimentally \cite{ilani_PRL_2000,ilani_2001, allison_PRL_2006} and  theoretically \cite{shi&xi_PRL_2002, asgari-tanatar_2002, oronzco_2003, fogler_PRB_2004}. Thus, the  calculations 
 Ref.~\cite{asgari-tanatar_2002} yield the overall $1/\kappa(r_s)$ dependence 
 similar to that observed   in the experiments 
 (Section~\ref{Field penetration}).

\subsubsection{Non-linear screening approach. Numerical modelling}
Shi and Xie \cite{shi&xi_PRL_2002} investigated spatial distribution of carrier
density and the compressibility of 2D electron systems
by using the local density approximation. A slowly varied disorder potential was applied to simulate the disorder effect. 
To investigate the density distribution of a disordered 2D
electron system,  within DFT, the total electron energy was calculated as
\begin{equation}
	E(n)=E_T(n)+ E_{ee}(n)+E_d(n) + Ex(n) + E_c(n)
\end{equation}

Here $E_T(n)$  is the functional of the kinetic energy, $E_{ee}(n)$ is the direct Coulomb energy due to charge inhomogeneity, $E_d(n)$ is the disorder potential energy, $E_x(n)$, and $E_c(n)$ are the exchange  and correlation energy, respectively. The ground state spatial density distribution was obtained by minimizing the total energy functional with respect to the density.  A slowly varied disorder potential was applied to simulate the disorder effect. Shi and Xie found that at low average densities electrons form a droplet state which is a coexistence phase of high- and low-density regions.        
In calculating  total exchange and correlation energy they used interpolated Tanatar and Ceperley QMC results for exchange and correlation energy density for the homogeneous 2DE system \cite{tanatar-ceperley_1989}.

\begin{figure}
	\includegraphics[width=200pt]{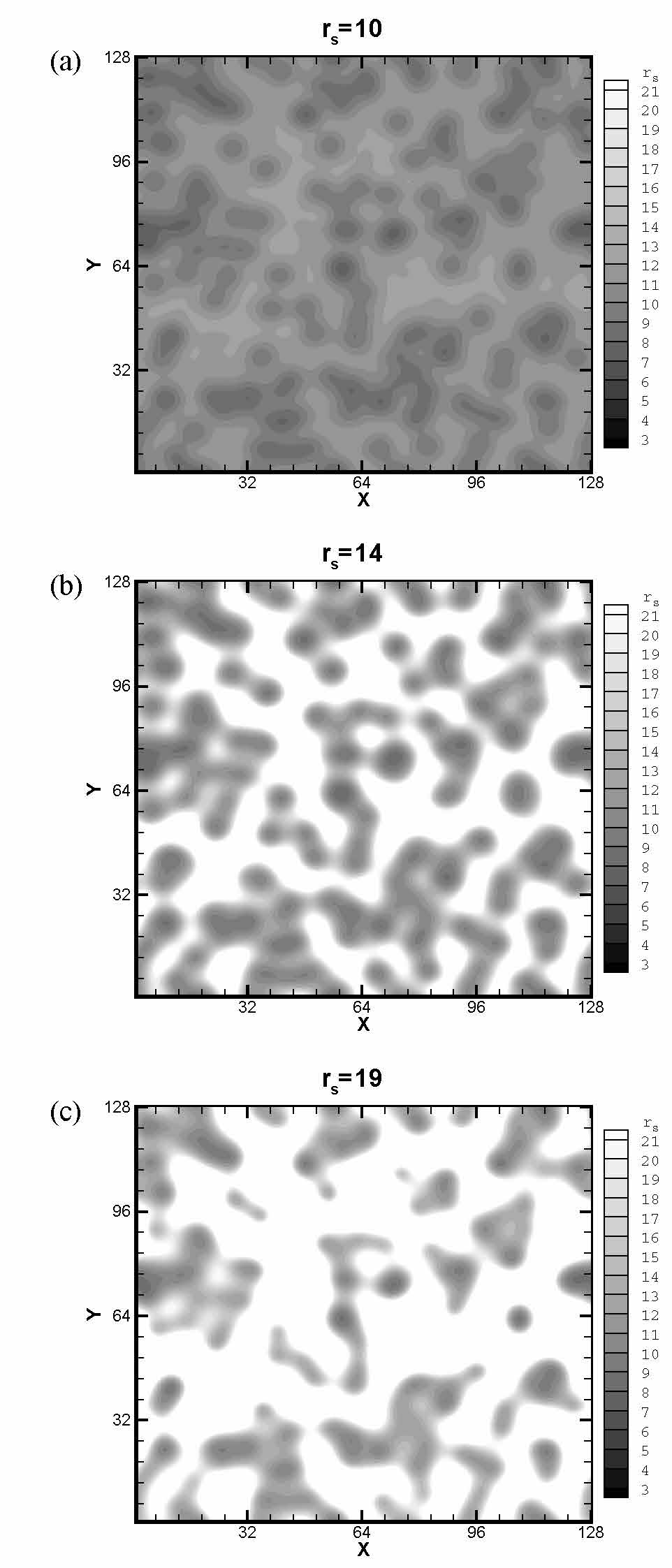}
	\caption{Spatial density distributions for various average densities.
		The contour plot shows  local density parameter $r_s =1/\sqrt{\pi n}$.
		The density in the white area decreases rapidly to zero. The size of the system is set as $L = 256a^*_B$. The disorder potential is generated by off-plane charge impurities with  $d = 10 a^*_B$, $n_i = 2.5 \times 10^{-3}/(a^*_B)^2$. Adapted from Ref.~\cite{shi&xi_PRL_2002}.
	}
	\label{shi-xie-fig1}
\end{figure}
It was found that the compressibility anomaly observed in 2D systems which accompanies the metal-insulator transition can be attributed to the formation of the droplet state due to a disorder effect at low carrier densities. 
Figure \ref{shi-xie-fig1} shows the density distribution of the system. It can be clearly seen that the electrons form
some high density regions, while the density of other  regions is essentially zero. Depending on the average
density of the system, the high-density regions may connect to each other ($r_s = 10$), or form some isolated regions
($r_s = 19$). There exists a certain density $r_s = 14$, where the connectivity of the high-density regions changes (a  percolation transition).

\begin{figure}
	\includegraphics[width=200pt]{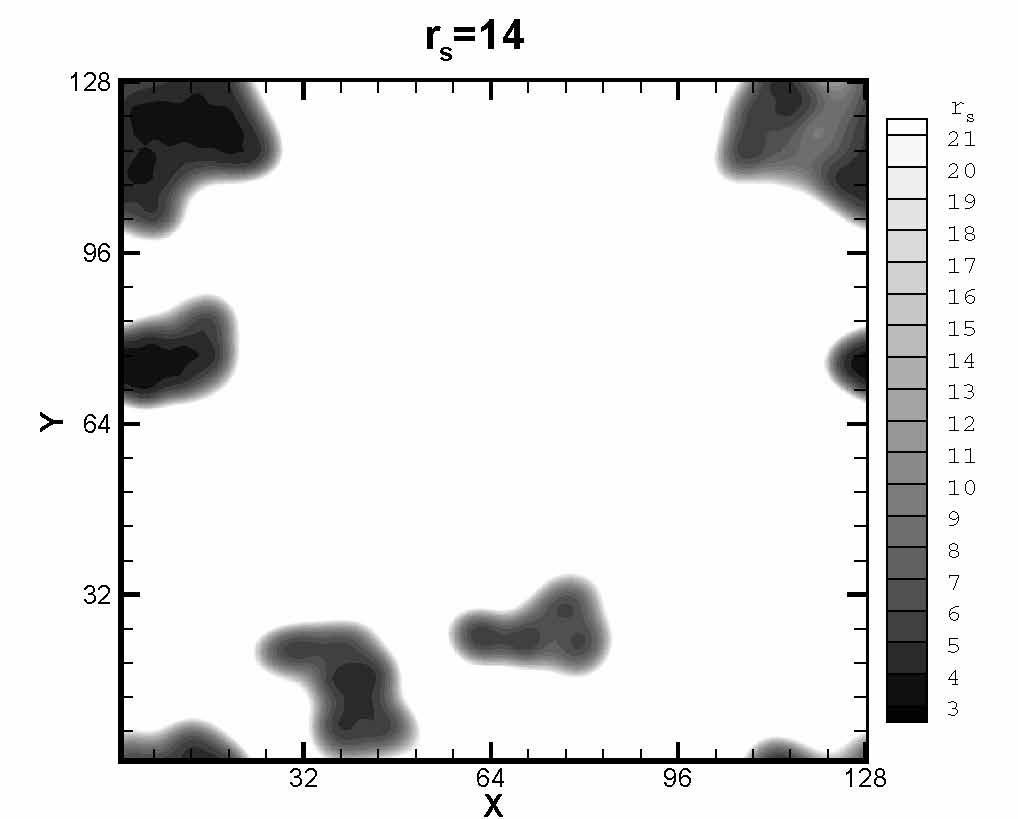}
	\caption{Spatial  density distribution for the free electron gas on
		the same disorder landscape as Fig.~\ref{shi-xie-fig1} at density $r_s = 14$. 
		Adapted from Ref.~\cite{shi&xi_PRL_2002}
	}
	\label{shi-xie-fig2}
\end{figure}

The e-e interaction is important for the conduction behavior of a dilute electron system in the
sense that it makes the density distribution more extended because of the Coulomb repulsion. 
Figure \ref{shi-xie-fig2}
shows the density distribution for the free electron gas with the same density as in Fig.~\ref{shi-xie-fig1} by turning off the electron-electron interaction. The system forms only some isolated high density regions at the most  disordered areas, while the density distribution of the corresponding interacting system (Fig.~\ref{shi-xie-fig1}b) is quite extensive at the same density. In other words,  at a given disorder strength, the critical density for the free electron gas is much higher than for its interacting analogue.

       \section{Compressibility of 2DE systems: Experimental studies}
\label{Experimental studies}

\subsection{Earlier capacitance measurements} 
        
In earlier experiments  \cite{krav_PLA_1989, krav_PLA_1990, krav_PRB_1990, krav_PRB_1993, goodall_1985}, information about compressibility  (or inverse density of states) was obtained from either capacitance measurements  or from  measurements of  the electrochemical potential variations 
versus density in quantizing magnetic field. In the former case, the capacitance was measured by AC-bridge in the frequency range 6-75\,Hz.
The measured capacitance  $C$  is considered to be a series connection of the geometric and ``quantum'' parts:
   \beq
    C^{-1} = C_0^{-1} + e^2 S\left(\frac{\partial n}{\partial \mu}\right)^{-1},
    \eeq
 where $C_0$ is the capacitance in the $\partial \mu/\partial n \rightarrow 0$ limit which does not depend on $B$, and $S$  is the 2DE layer area. $C_0$ may be estimated as follows:
       $$
        C_0^{-1}= C^{-1}|_{B=0} - \left(e^2 S D_0   \right)^{-1},
        $$
 where the density of states $D_0=\partial n/\partial\mu|_{B=0} = 8.37\times 10^{14}(m^*/m_e)$cm$^{-2}$eV$^{-1}$ \cite{ando_review}.
 In order to separate the second ``quantum''  part from the geometric capacitance, the data taken  in zero field was subtracted from that in magnetic field.       Correspondingly, the zero field behavior of the compressibility remained inaccessible.

\begin{figure}
\includegraphics[width=230pt, height=130pt]{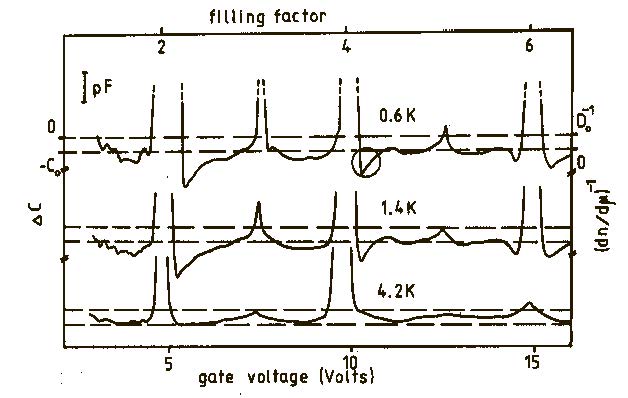}
	\caption{$\Delta C$ dependences on $V_g$ (proportional to the carrier density) for three temperatures and at field 11.7\,Tesla. The upper horizontal axis shows the Landau level filling factors.
		Adapted from Ref.~\cite{krav_PLA_1989}	
	}
	\label{krav_1989-fig1}
\end{figure}

Figure \ref{krav_1989-fig1} represents the difference $\Delta C (n)$ of the two  dependences 
        $
        \Delta C = - \left(C - C|_{B=0}  \right)
        $
measured  with Si-MOSFET sample as a function of carrier density at three temperatures and in a fixed magnetic field of 11.7\,Tesla.
The measured difference equals to
        $$
        \Delta C \approx \left( C^2/e^2 S \right) \left[ (\partial n/\partial \mu)^{-1} - D_0^{-1}\right]
        $$
        
The obtained dependence at $T=4.2$\,K  agrees with earlier capacitance measurements \cite{smith_1986, smith_1985, klitzing_1986}. In particular, the
$\left(\partial n/\partial\mu \right)^{-1}$ values at half-integer fillings are less than $D_0^{-1}$ but positive.
However, lowering the temperature to 1.4\,K and to 0.6\,K leads to the appearance of regions of filling factors where $(\partial n/\partial \mu)^{-1}$ gets negative. The appearance of these dips in the vicinity but somewhat away of the integer filling factors was exactly the feature predicted by Efros \cite{efros_SSC_1988}. Indeed, in his language, the total electron energy acquires a term $E_{ee}$ in addition to the single-particle term $E_{1p}$. Therefore, the inverse total density of states may be written as
  $$
        D^{-1} =D^{-1}_{1p} + G^{-1}_{ee}.
        $$
        $G_{ee}$ was evaluated by Efros \cite{efros_SSC_1988} as follows:
        \bea
        G_{ee} &=& -(e^2\alpha/\varepsilon)\left(\{\nu\}n_B\right)^{3/2}, \qquad \{\nu\} \leq \frac{1}{2}, \nonumber\\
        &=& - (e^2\alpha/\varepsilon)\left[\left(1-\{\nu\}\right)n_0^{3/2}\right], \qquad \{\nu\} \geq \frac{1}{2} \nonumber,
        \eea
 where $\nu=n/n_B$ is the Landau levels filling factor, $n_B=1/(2\pi l_B^2)$ -- the Landau level degeneracy, 
 $\{\nu\}$ - fractional part of the filling factor $\{\nu\}= \nu- int(\nu)$, and $\alpha$ is  a dimensionless constant ($=2$ for the classical Coulomb interaction).
      
 The inverse thermodynamic density of states, correspondingly, is equal to \cite{efros_SSC_1988}:
        \bea
        D^{-1}_{ee} &=& -\left(\frac{3\alpha e^2}{4\pi\varepsilon}\right)\left(\{\nu\}n_B\right)^{-1/2}, \quad \{\nu\} \leq \frac{1}{2}, \nonumber\\
        &=& -\left(\frac{3\alpha e^2}{4\pi\varepsilon}\right)\left[(1-\{\nu\})n_B \right]^{-1/2},  \{\nu\} > \frac{1}{2} 
        \label{efros-eq2}
        \eea

The negative compressibility signals a tendency of the 2DE system to break the homogeneous state.
On the other hand, the stability of the entire 2D system with a negative compressibility 
is achieved in physical systems by the neutralizing background. For the gated 2D system, the stability condition 
was analyzed in Ref.\cite{nielson_PhysB_1993}.
           
\subsection{Field penetration measurements}
\label{Field penetration}
In the conventional capacitance technique, the capacitance between the 2D gas and a metal gate electrode is
measured. The dominance of a large geometric term in the measured capacitance
essentially forces one to vary some other parameter, like magnetic field 
\cite{krav_PLA_1989, krav_PLA_1990, krav_PRB_1990, krav_PRB_1993}, temperature \cite{tupikov_JETPL_2015, tupikov_NatCom_2015}, etc., and then subtract off a large, and constant, offset in order to uncover the quantum term. There are three major drawbacks to this technique. First, the geometric term is usually not accurately known and therefore the subtraction is uncertain. The geometric term produces
a second difficulty as well: it may not actually remain constant as the external parameter, e.g. magnetic field is changed. Thirdly, the slowly-decaying eddy currents  excited in 2DE system  by ac-modulation of the carrier density \cite{pudalov-nonstationary} impede capacitance measurements at low temperatures in quantizing magnetic field.
       
The ``floating gate'' technique elaborated in Ref. \cite{pudalov_JETP_1985} does not require field or density modulation and therefore can be used for electrochemical potential measurements even in the QHE regime \cite{pudalov_JETPL_1986, pudalov_UFN_2021}. 

\begin{figure}
\includegraphics[width=200pt]{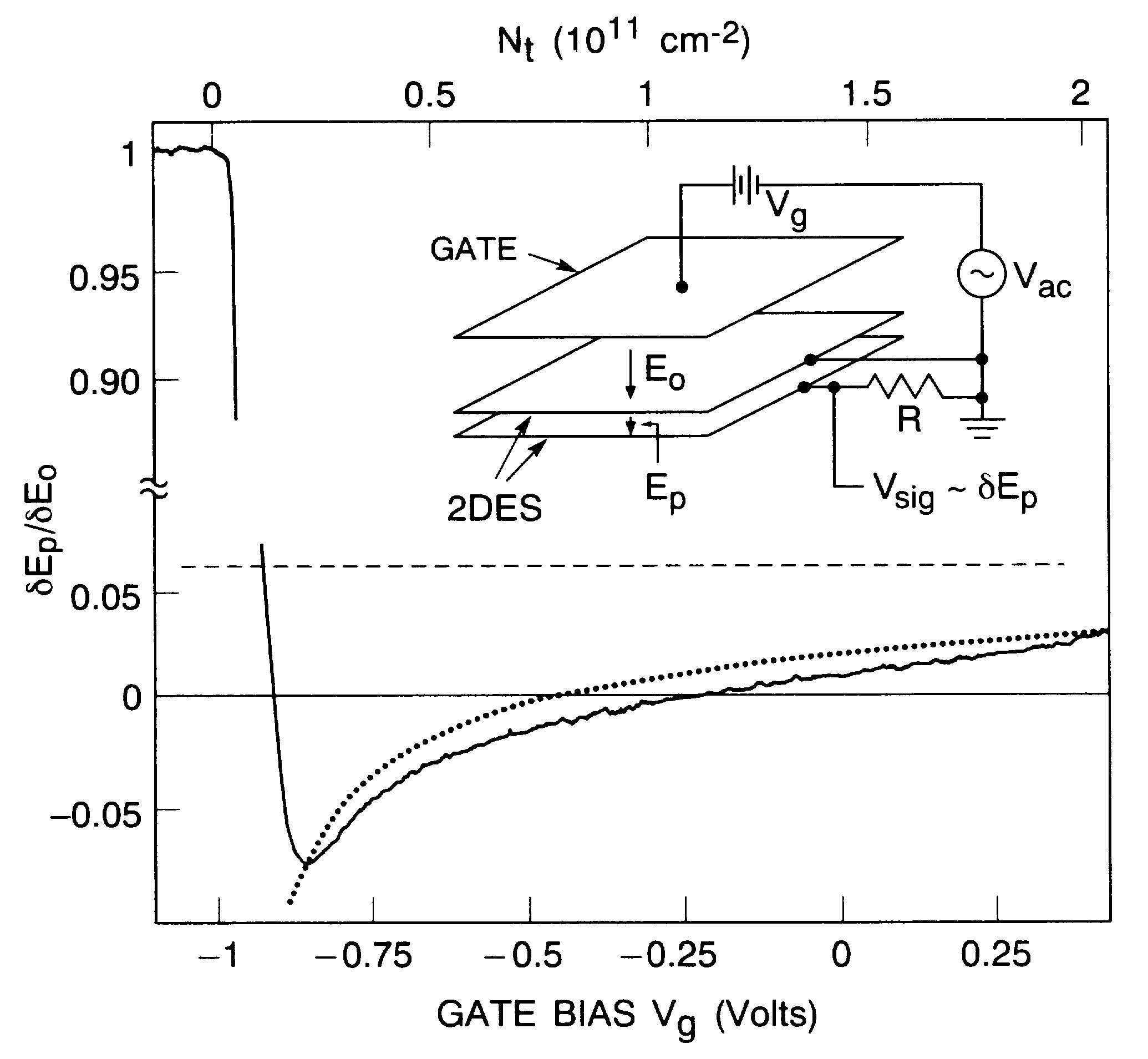}
\caption{Normalized penetrating field $\delta E_p/\delta E_0$ versus gate voltage at zero magnetic field and $T=1.2$K. Dotted curve  is calculated using Tanatar and Ceperley's compressibility \cite{tanatar-ceperley_1989}. Upper axis gives carrier density of the top 2DE system. Dashed horizontal line - noninteracting case. Inset - experimental set-up.
Adapted from Ref.~\cite{eisenstein_1992}.
}
\label{eisenstein_1992-fig1}
\end{figure}

The alternative field penetration technique introduced by Eisenstein \cite{eisenstein_1992, eisenstein_1994}  provides automatically subtracting the geometric term. This is achieved by use of a double layer 2D system and measuring the fraction of the ac electric field $\delta E_0$ which penetrates one layer and is detected by the second. 
The inset to Fig.~\ref{eisenstein_1992-fig1} shows a schematic set-up. The ac  field $\delta E_p$   penetrates through the upper layer and causes current flow  through the external impedance $Z$, thus generating a detectable voltage $V_{sig}$.       

Lower part of Fig.~\ref{eisenstein_1992-fig1} shows a trace of the measured ac-current (proportional to penetration field) as a function of the gate voltage or electron density of the upper 2D layer. The penetration field measures the screening ability of the electrons which is inversely proportional to $\kappa$. The main advantage of 
this experimental approach 
is that it provides direct access to $\partial \mu/\partial N$ for the top 2DE layer without any offset signals (related with geometric capacitance contribution). Dultz and Jiang \cite{dultz&jiang_PRL_2000} have extended the field penetration method to a more conventional heterostructure with only a single layer of carriers. 

Were the 2DE system noninteracting, the penetration to the bottom layer would be a few percent and would be
positive. This result is qualitatively altered by e-e interactions that make the
observed differential penetration negative. The minimum in the oinverse compressibility was already detected in the measurements by Eisenstein \cite{eisenstein_1994}.

This minimum attracted much interest when Dultz and Jiang \cite{dultz&jiang_PRL_2000} reported that in some samples the minimum virtually coincides with the metal-insulator transition  in transport.
They found \cite{dultz&jiang_PRL_2000} that the negative $1/\kappa$ at low densities reaches a
minimum value at a certain density, and then increases
dramatically with further decreasing $n$. This coincedence was initially considered as a thermodynamic signature of an interaction-driven phase transition 
\cite{chakravarty_PhilMag_1999, si-varma_1998}. 
However, later on Allison et al. \cite{allison_PRL_2006}  measured 
simultaneously the compressibility, capacitance and resistivity in the vicinity of the metal-insulator transition
with different samples. It was shown that the coincedence of the two effects in some samples, the inverse compressibility minimum and the sign change in transport $d\rho/dT$, is accidental.

       \subsection{Local compressibility measurements}
Ilani et al. \cite{ilani_PRL_2000,ilani_2001}  have performed  local  study of $\partial\mu/\partial n$ and expanded them into the low density regime, across transition to the localized state. Their measurements utilized  single
electron transistors (SET), located directly above a two-dimensional hole gas (2DHG) of inverted back-gated GaAs/AlGaAs structures. This technique allowed  to probe the local behavior of $\partial\mu/\partial n$ as well as its spatial variations.  At equilibrium,
the Fermi energy is constant across the sample, and therefore a change in $\mu(n)$ induces a change in the electrostatic potential, which is readily deduced by measuring the change in current through the SET.  The spatial resolution, determined from the size of the SET and its distance from the 2DHG, is $0.1 \times 0.5\mu$m$^2$.

\begin{figure}
\includegraphics[width=170pt]{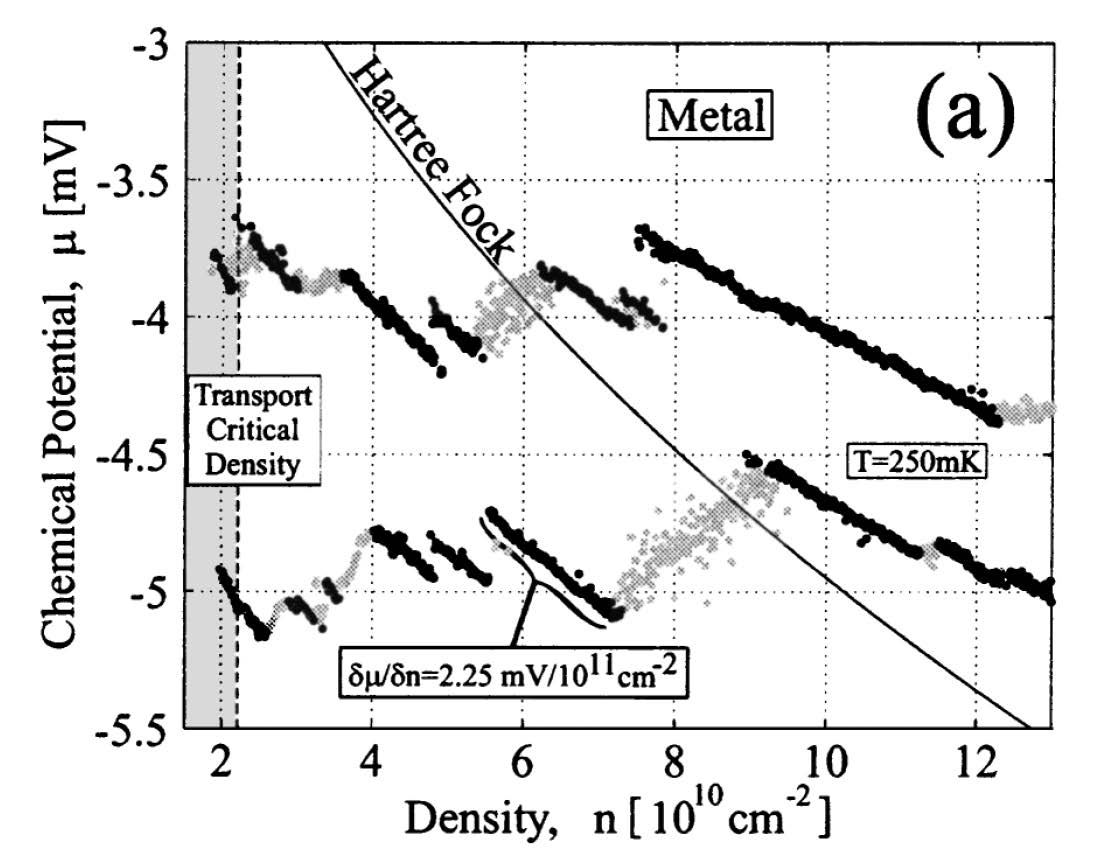}
	\label{ilani-fig3a}
\includegraphics[width=170pt]{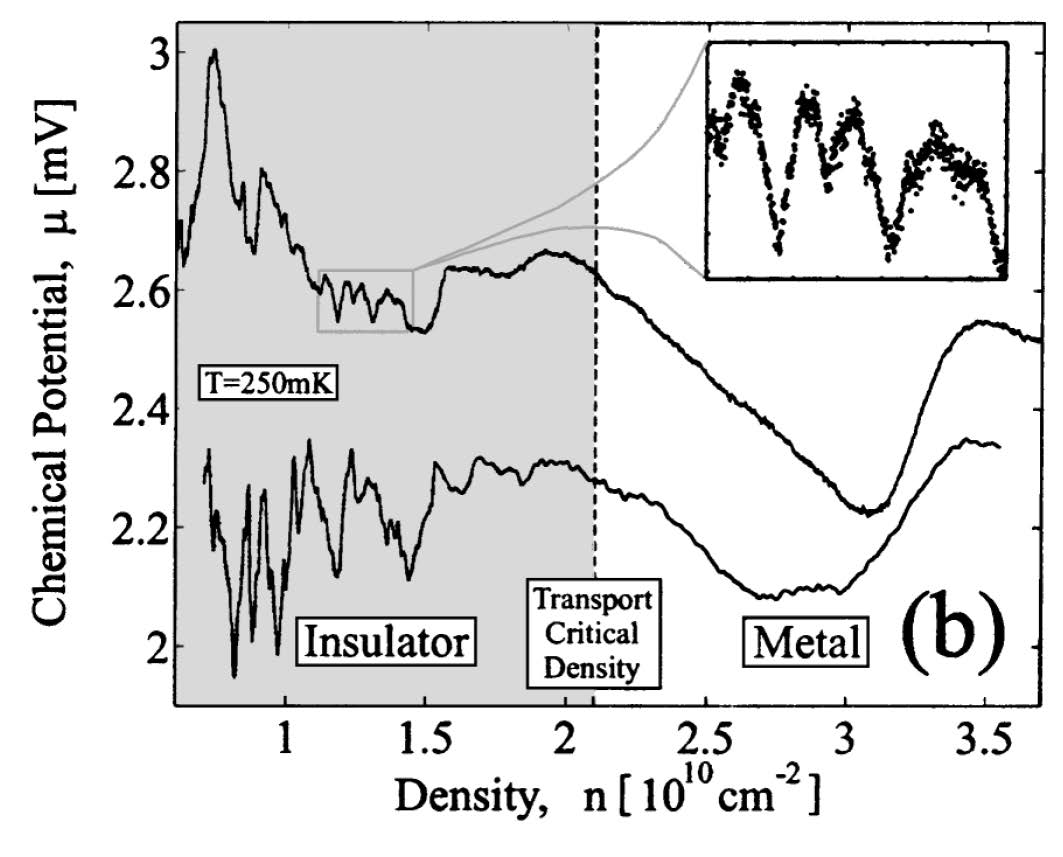}
	\caption{(a) Measured $\mu(n)$ in the metallic regime (dots) together with the HF theory [Eq. (1)] for a clean system (solid line). The measured
		negative slopes are highlighted (dark symbols) to demonstrate their resemblance to the HF model. (b) Measured $\mu(n)$ across
		the MIT and in the insulating region. Inset: A closer look at the data in the insulating regime. Each slope is composed of many
		data points allowing an accurate determination of the slopes. Adapted from Ref.\cite{ilani_PRL_2000}
	}
	\label{ilani-fig3}
\end{figure}

Instead of the anticipated
monotonic dependence, local $\mu(n)$ exhibits a rich structure of oscillations. 
In the high density ``metallic'' regime, Ilani et al. observed long sawtooth oscillations.
Superimposed on them and starting in close proximity to the onset of  
the localized state, a new set of rapid oscillations emerges (Fig. \ref{ilani-fig3}b). Their typical period is an order of magnitude smaller, 
and the amplitude grows continuously from the point of appearance to lower densities.  
All the oscillations, including the fine structure seen on the left side of
Fig.~\ref{ilani-fig3}b, were reproducible and did not 
exhibit hysteresis or sweep rate dependence.

The sawtooth profile is reminiscent of the  electrochemical potential behavior  for a quantum 
dot as a function of the number of  electrons \cite{kouwenhoven_review} and, hence,
suggests the existence of discrete charging events.
Thus, the measured $\mu$ of the 2DHG varies undisturbed along the segments with negative slopes, until a certain bias  between the 2DHG and the SET  makes recharging of an intermediate localized state energetically favorable. This causes a 
sharp drop in the electrostatic potential, after which $\mu$ continues to vary smoothly 
until the next screening event occurs.

\begin{figure}[ht]
\includegraphics[width=200pt]{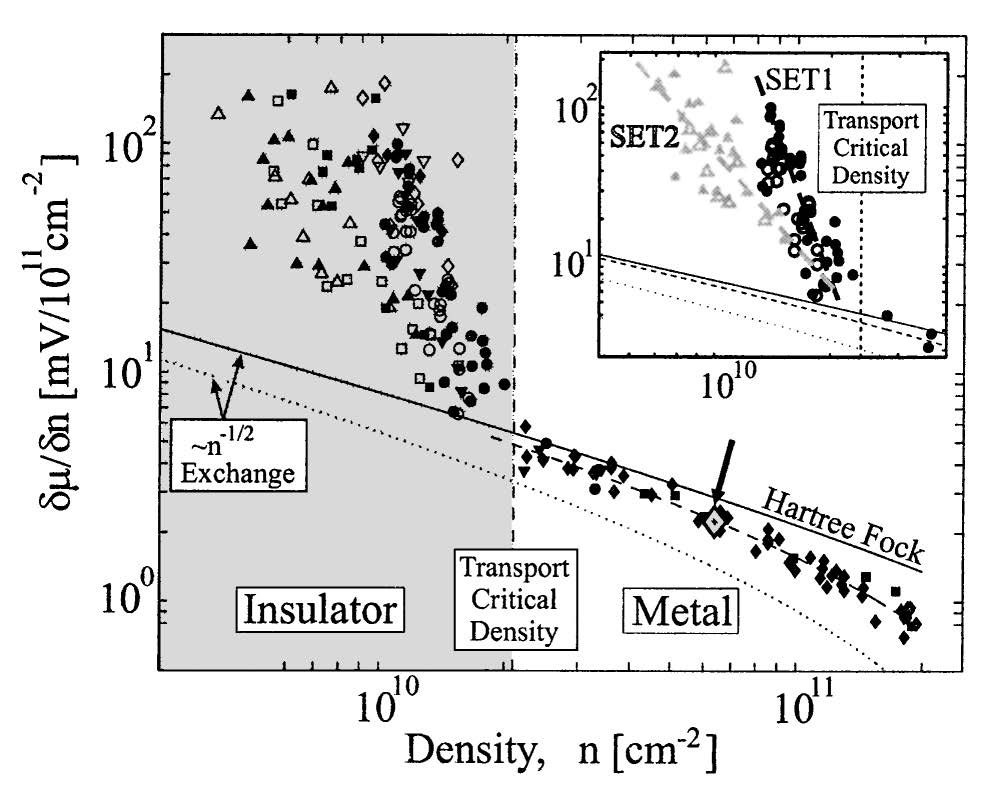}
	\caption{$|d\mu/dn|$ collected from several  SETs on several samples. In the insulating regime,
		the $|d\mu/dn|$ magnitude includes both negative and positive slopes. Each point corresponds to a well-defined
		segment in the $\mu(n)$ trace. The point marked by an arrow corresponds to the marked segment in Fig.~\ref{ilani-fig3}a. Inset: Results
		from two SETs on the same device  demonstrating the spatial dependence of $|d\mu/dn|$ in the insulating side. Adapted from Ref.~\cite{ilani_PRL_2000}
	}
	\label{ilani-fig4}
\end{figure}

Ilani et al.  reconstructed the basic  $\mu(n)$  dependence
by assembling  the  undisturbed segments together;  the results are shown in  Fig.~\ref{ilani-fig4}, 
for five different SETs placed apart from each other.
In the ``metallic'' regime (high density) all the data collapse onto a single curve  in Fig.~\ref{ilani-fig4}, rather close to  the HF model prediction.

In the insulating phase, assuming that the new set of oscillations is caused by the same
mechanism (screening by traps), the authors  extracted the slopes and added them to the same plot of
$\partial \mu/\partial n$ (see Fig.~\ref{ilani-fig4}). Unlike in the metallic phase where the system clearly has a negative $\partial \mu/\partial n$,
in the insulating phase the sign of the compressibility is not known a priori. Therefore, in Fig.~\ref{ilani-fig4} the absolute value is shown  of both negative and positive slopes, 
all of which deviate considerably from the expected $n^{-1/2}$ power law. The deviation
becomes greater than an order of magnitude at the lowest density and indicates  a change in the screening 
properties of the 2DHG at the transition to the localized state. The 
fluctuations in the slopes, observed on the insulating side, are  reproducible and suggest that mesoscopic effects are present.  Furthermore, the average behavior of $\partial\mu/\partial n$ in this fluctuating regime is  position
dependent (see the inset in Fig.~\ref{ilani-fig4}).
Such dependence on position indicates that once the system crosses into the insulating
phase it becomes spatially inhomogeneous.

 The  local measurements \cite{ilani_2001} emphasize the important role of charged traps in the  ground state thermodynamics of the 2D system. It might have a direct relationship  to the models where the 2D gas and charge traps  coexist in equilibrium, particularly \cite{altshuler-maslov_1999, pudalov-nonlinear_PRB_2021}.
To summarize shortly the results of local measurements ~\cite{ilani_PRL_2000}, it was  found that the behavior of $\partial \mu/\partial n$ in 
the metallic phase on average follows the HF model suggesting the 2D system to be almost spatially homogeneous. In contrast, the insulating phase is found to be spatially inhomogeneous.

 \section{Phase separation effects revealed in thermodynamics and transport}  

    \subsection{Evidence of the ``spin-droplet'' state in thermodynamic spin magnetization}
    \label{spin droplets}
The method of $\partial\mu/\partial B$ thermodynamic measurements was introduced and substantiated  in 
Refs.~\cite{prus_PRB_2003, reznikov_JETPL_2010}.
 To probe purely spin susceptibility, free of  orbital contribution, measurements in 
Refs.~\cite{prus_PRB_2003, reznikov_JETPL_2010} were performed in  magnetic field 
$B_\parallel$ aligned strictly parallel to the 2D plane.     
In this technique, the  applied magnetic field is modulated with a small amplitude $\delta B$ and the excited recharging current of the Si-MOS straucture \cite{reznikov_JETPL_2010} is measured: $\delta I= 
  [i\omega C_0 \delta B/e] (\partial\mu/\partial B)$.
Here $C_0$ is the known capacitance of the ``gate - 2D layer'' structure.
From the measured recharging current the quantity $\partial\mu/\partial B$ is found  and, 
due to the Maxwell relation, 
directly yields the magnetization per electron $\partial M/\partial n$.

\begin{figure}[ht]
	\includegraphics[width=190pt]{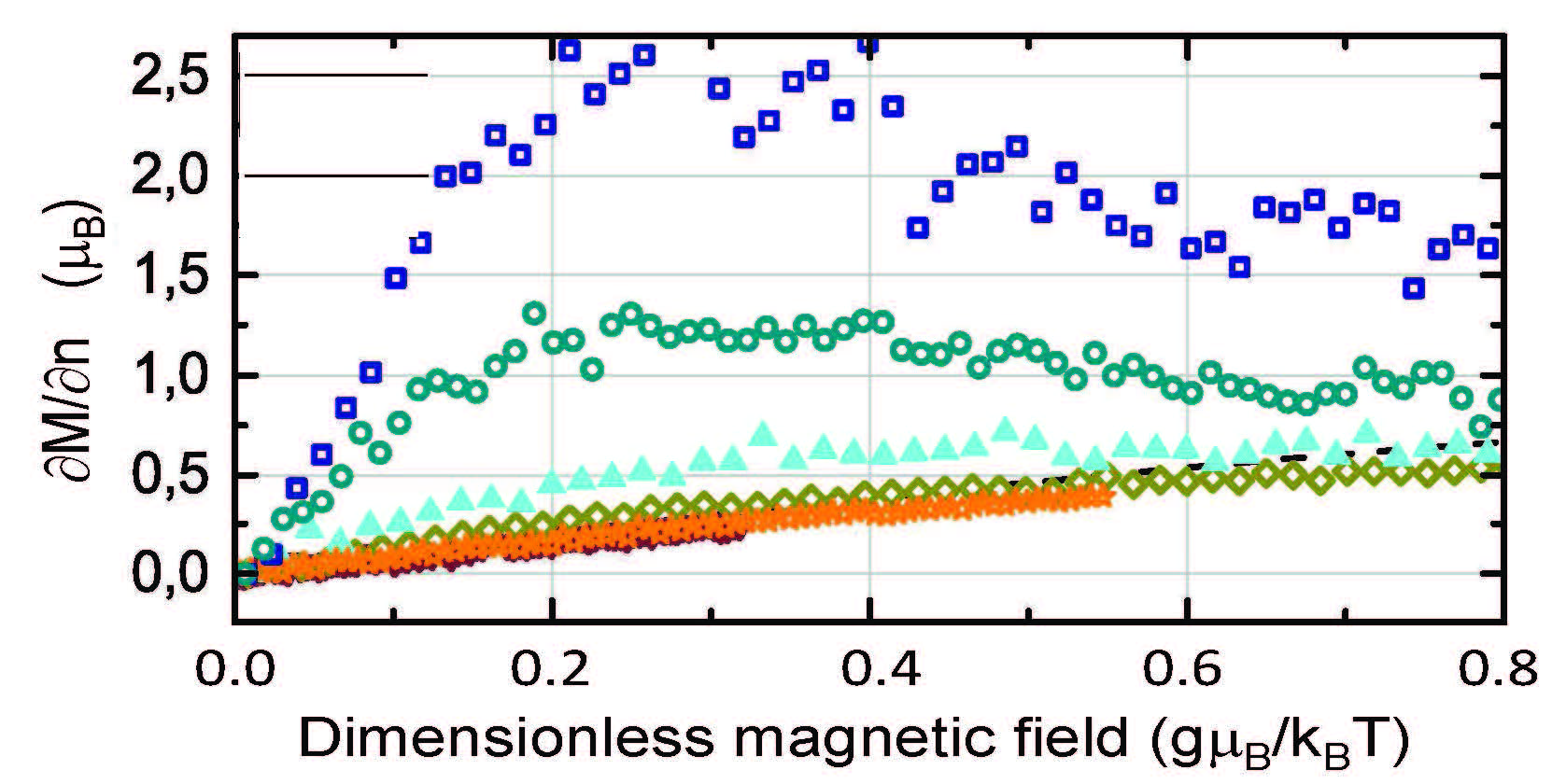}
	\caption{Magnetization per electron $\partial M/\partial n$ in weak field  plotted versus the normalized
	magnetic field 
	for a carrier density of $5\times 10^{10}$cm$^{-2}$ at
several temperatures ($T = 0.8$, 1.2, 1.8, 4.2, 7, 10, 24 K, from top down).
	 Adapted from Ref.~\cite{teneh_PRL_2012}
	}
	\label{teneh_PRL_2012}
\end{figure}

In order to explore interaction effects the measurements in \cite{teneh_PRL_2012} 
were performed in weak fields less than temperuture, $g\mu_B B \leq k_B T$.    
In Figure \ref{teneh_PRL_2012} one can see that at low densities $\partial M/\partial n$ 
becomes positive and in all cases
is much greater than expected for the Pauli spin susceptibility.
When the field increases (while still being smaller than the
temperature, $g\mu_BB < k_BT$), $\partial M/\partial n$ sharply increases and
exceeds the Bohr magneton by more than a factor of two at
low temperatures (Fig.~\ref{teneh_PRL_2012}).
Such behavior of $\partial M/\partial n$ is reminiscent of the
dependence anticipated for free spins \cite{teneh_PRL_2012}. However, the fact that  $\partial M/\partial n$ exceeds the Bohr magneton points to a ferromagnetic ordering of the electron
spins. The magnetization curves  $\partial M/\partial n$ (Fig.~\ref{teneh_PRL_2012}) saturate
in the field $b=\mu_B B/(k_BT) \sim 0.25$, signaling that the particles which
respond to the field modulation have spins $1/(2b) \approx 2$,
rather than 1/2.
This result is the ``smoking-gun-evidence'' of the emergence of a
two-phase state in the 2D system consisting of a paramagnetic
Fermi liquid and ferromagnetic domains (called ``spin droplets'') with the total spin $\sim 2$,
 comprising, respectively, four or more electrons.
 
 The existence of the two-phase state is not a unique property of the low-density state solely,
 the spin droplets were detected in Ref. \cite{teneh_PRL_2012} in the wide range of densities, up to  $n \approx 2\times 10^{11}$cm$^{-2}$ that is twice the critical density of the onset of the insulating state, i.e. in   
 the  regime of high metallic metallic conduction $\sigma \approx 80 e^2/h$.

    \subsection{Phase separation effects in charge transport}
    
In Ref.~\cite{morgun_PRB_2016}, several features have been revealed in magnetotransport, zero-field transport, and thermodynamic spin magnetization for a 2D correlated electron system. These features  have been associated 
with the two-phase state. More specifically:\\ 
(i) in magnetoconductivity   the novel regime of magnetoconductance 
sets above a density-dependent temperature $T_{\rm kink}(n)$.\\
  (ii) In the temperature dependence of zero-field
resistivity  an inflection point is observed at about the same temperature $T_{\rm infl}(n) \approx T^*$.\\
(iii) In thermodynamic  magnetization the weak-field spin susceptibility per electron, $\partial \chi /\partial n \equiv \partial^2 M/\partial n$
changes sign at $T_{dM/dn}(n) \approx T^*$.

\begin{figure}[ht]
	\includegraphics[width=240pt]{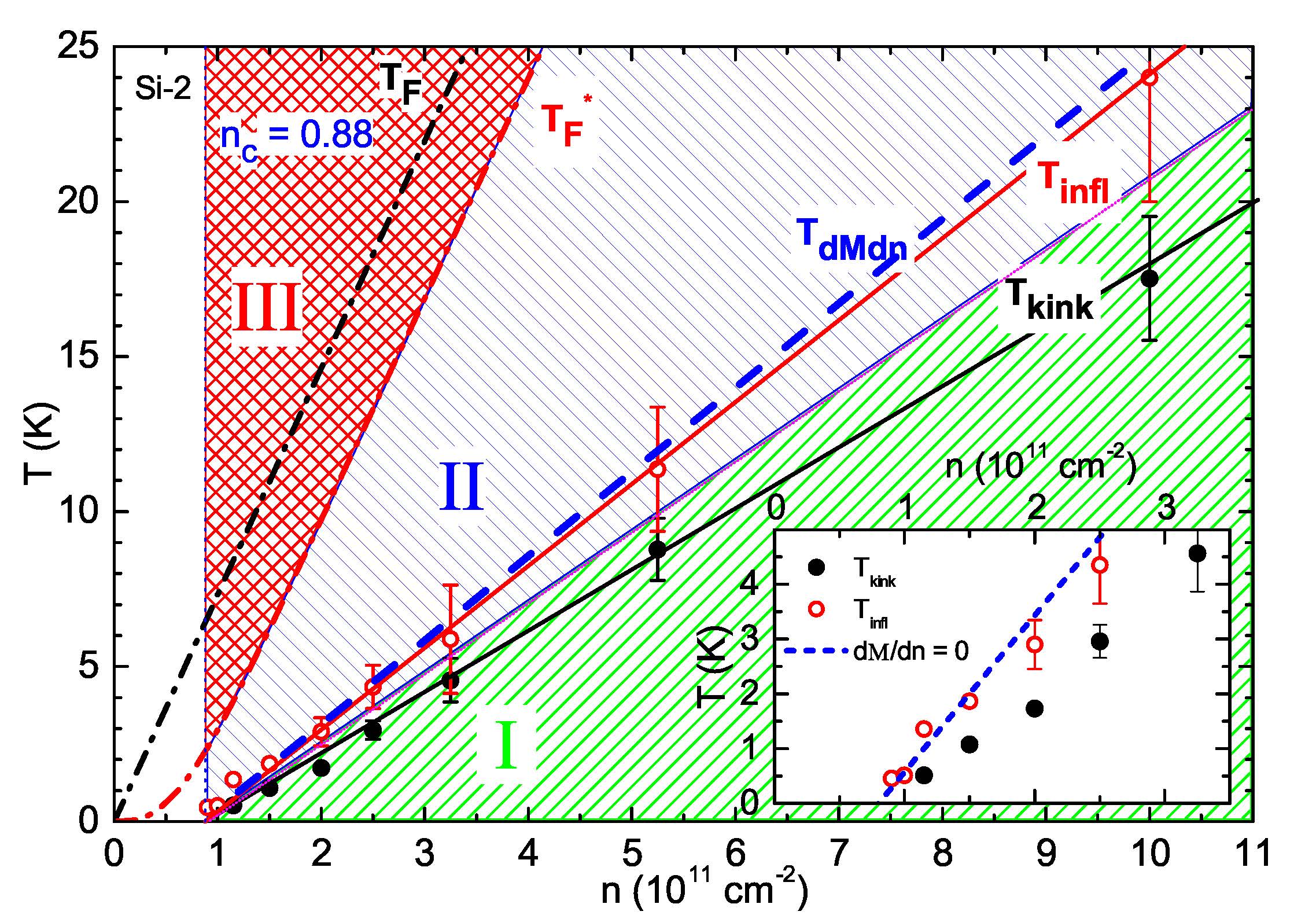}
	\caption{Empirical phase diagram of the 2DE system. Dashed
		areas are (I) the ballistic interaction regime and (II) the anomalous
		magnetoconductance regime. Hatched area (III) is the nondegenerate regime, the blank
		area at $n < nc$ is the localized phase. Full dots: the kink temperature
		Tkink; open dots: the inflection point $T_{infl}$.  Dash-dotted
		curves show the calculated bare ($T_F$) and the renormalized ($T^*_F$ ) Fermi
		temperatures. The insert blows up the low-density region; the dashed
		line is $T_{dM/dn}$ \cite{teneh_PRL_2012}. Adapted from \cite{morgun_PRB_2016}.
	}
	\label{fig:phase diagram}
\end{figure}

All three notable  temperatures, $T_{\rm kink}$, $T_{\rm infl}$, and $T_{d M/ d n}$ 
are close to each other {\cb (see Fig.~\ref{fig:phase diagram})},  and are intrinsic to strongly correlated regime solely.
It is shown below that these features can be described within the framework of the phase separation approach.

\subsubsection{Magnetotransport in the in-plane field}

Regarding the zero field transport and magnetotransport, their features require more detailed explanation.

In the conventional  theory of interaction corrections (IC) \cite{ZNA_BPar_2001}, 
the lowest order  variations of the magnetoconductivity (MC)  
with weak in-plane field  $g\mu_B B< k_BT \ll
k_B T_F$ at a fixed temperature $T$
are  parabolic. This is clear  from symmetry arguments, and also follows from the IC theory and the screening theory \cite{gold-dolgopolov_PRB_1986, g-d-MR_JETPL_2000}.
\beq
\sigma = \sigma_0 - a_\sigma  B^2 + {\cal{O}}\left(B^2\right); \quad \rho = \rho_0 + a_\rho B^2 + {\cal{O}}(B^2),
	\label{eq:sigma}
\eeq
where   by definition
$$
	a_\sigma \equiv  \left. -\frac{1}{2} \frac{\partial^2\sigma}{\partial B^2}\right|_{B=0} =  \frac{1}{2 \rho^2} \frac{\partial^2 \rho}{\partial B^2}; \quad 
	a_\rho \equiv  \left. \frac{1}{2} \partial^2\rho/\partial B^2 \right|_{B=0} \nonumber.
$$

In Ref.~\cite{morgun_PRB_2016} the in-plane field MC was studied in detail 
and quantified in terms of the prefactor $a_\sigma(T,n)$.
Within the IC theory, the $\sigma(T,B)$
variation in the 2DE system is described by the sum of the interference correction
and e-e interaction corrections \cite{ZNA-R(T)_PRB_2001, ZNA_BPar_2001}
$$
\Delta\sigma(T) \approx \Delta\sigma_C(T,B) +n_T(B)\Delta\sigma_T(T,B) +O\left(\frac{1}{k_Fl}\right).
$$
Here the first term combines single-particle interference and  interaction corrections in the singlet channel,
and the second term is the interaction corrections in the triplet channels, $k_Fl  \gg 1$ is the dimensionless conductivity.
Within the same approach, MC in the weak in-plane field originates from  the field dependence of the
effective number of  triplet channels $n_T(B)$, that in its turn is due to the Zeeman splitting  \cite{ZNA_BPar_2001}.

As a result,  the first order interaction corrections to MC  in the diffusive  and ballistic interaction regime
$\Delta\sigma 
\equiv \sigma(T,B)-\sigma(T,0)$ may be written in terms of $a_\sigma$ as follows \cite{ZNA_BPar_2001}:
\begin{eqnarray}
	a_\sigma(T) \propto \left\{
	\begin{array}{cl}
		&(1/T)^2, \quad   T\tau \ll 1\\
		&(1/T),    \quad  T\tau \gg 1.\\
	\end{array}
	\right.
	\label{eq:MC_IC}
\end{eqnarray}
Their explicit expressions  are given in Ref.~\cite{ZNA_BPar_2001}.

\begin{figure}[ht]
\includegraphics[width=240pt]{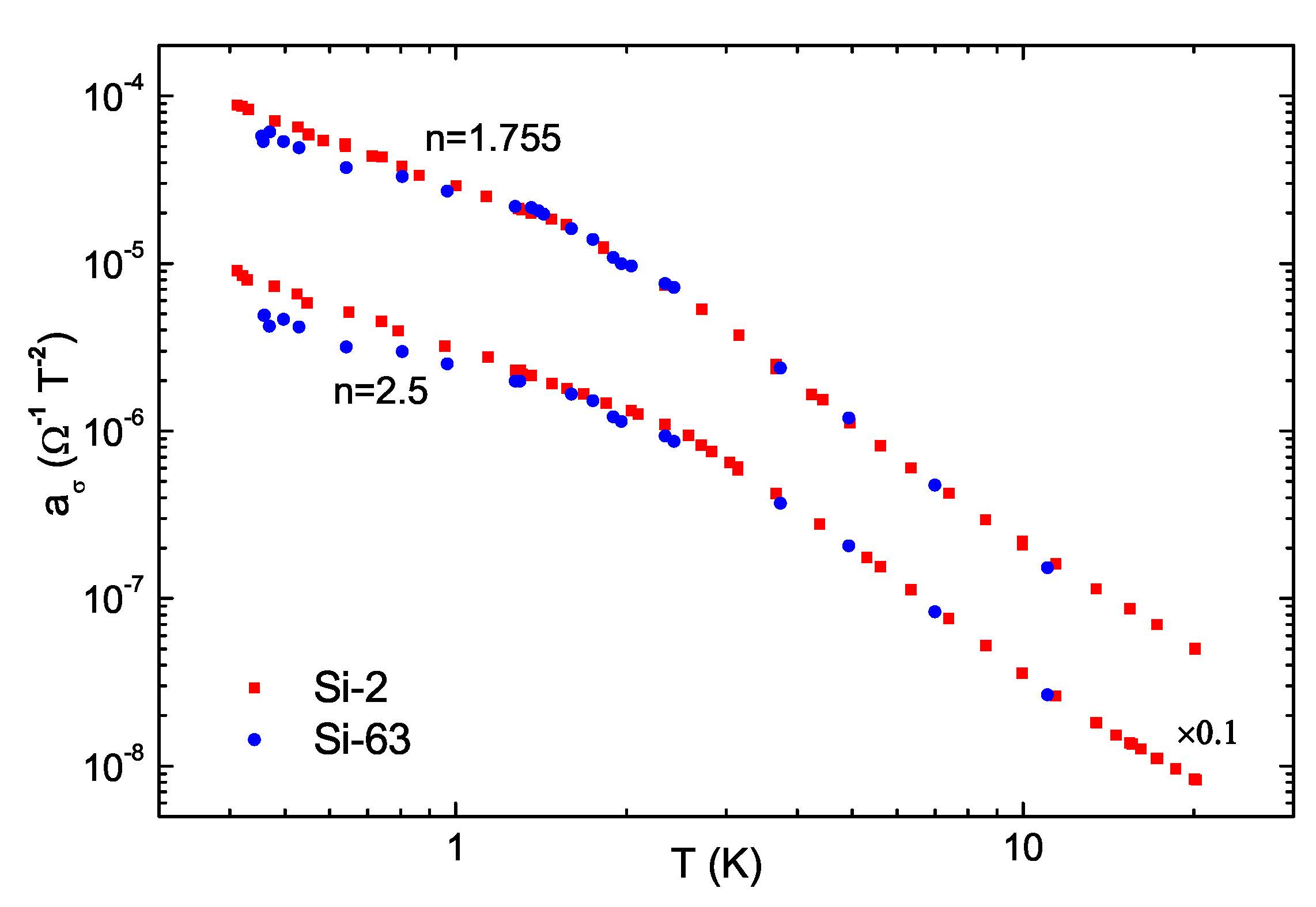}
	\caption{Comparison of the temperature dependences of the  prefactors   $a_\sigma(T)$  for two samples
		Si2 and Si-63, and for two density values (in units of $10^{11}$cm$^{-2}$.  For clarity, the  curves are  
		scaled by the factors shown next to each curve.}
	\label{fig:a_sigma-SiX}
\end{figure}

Thus, according to the IC theory predictions, as temperature {\em increases}, the MC  should cross over from 
$(1/T^2)$ to $(1/T)$ temperature dependence. This theory prediction is confirmed in measurements with low-mobility 
(high density, weak interactions) Si-MOS samples \cite{morgun_PRB_2016}.
In contrast, for high-mobility (lower densities, strongly interacting regime) structures,  as Fig.~\ref{fig:a_sigma-SiX} shows, with temperature {\em increasing} $a_\sigma(T)$ crosses over from the conventional ballistic-type $-(B^2/T)$ to the unomalous $-(B^2/T^2)$ dependence. Despite the absence of overheating of electrons \cite{prus_PRL_2002}, the diffusion regime of MC in the high mobility structures is not observed down to $T=0.3$K.

One can see from Fig.~\ref{fig:a_sigma-SiX}, that the ballistic-type behavior $\propto T^{-1}$ 
extends up to temperatures 1.5-2\,K (which are a factor of 10 higher than the estimated diffusive/ballistic border $T_{\rm db}\approx 0.2$K) \cite{morgun_PRB_2016}, then it sharply changes to  the novel  
dependence, $a_\sigma(T)
\propto T^{-2}$, making the overall picture clearly inconsistent with theory predictions, Eq.~(\ref{eq:MC_IC}). 
The crossover in Fig.~\ref{fig:a_sigma-SiX} occurs rather sharply,  as a kink on the double-log scale.
The kink and the overall type of behavior were  observed in the wide range of densities and were qualitatively similar
for several studied high mobility samples.   The next section shows that the observed effect in the in-plane magnetic field is associated with the onset of the two-phase state.

\subsubsection{Phase separation effects in oscillatory magnetotransport}

Measurements of the oscillatory magnetoresistance in high mobility  Si-MOS structures  in weak perpendicular magnetic fields were performed in Ref.~\cite{pudalov-gersh_JETPL_2020}. It was found 
that the quantum oscillations in 2D electron systems are observed {\em down 
to the critical carrier density} $n_c$ of the transition to strongly localized state. For such low densities, the oscillations exhibit an anticipated period, phase, and amplitude, even though the conductivity becomes essentially less than $e^2/h$, and, hence, the mean free path becomes less than the Fermi wavelength $\lambda_F$. It was concluded that this apparent contradiction with the Ioffe - Regel criterion for diffusive transport
is caused   by the emergence of an  inhomogeneous state of the 2D system, in which the regions of diffusive and hopping conduction are spatially separated.

The existence of quantum resistivity oscillations down to the critical electron density provides an evidence
for emerging inhomogeneity of the 2D system.
As density approaches $n_c$, the ``global'' resistivity, calculated under assumption of a uniform current flow, becomes
much greater than the ``local'' resistivity of the spatial areas, which contribute primarily to the oscillations
amplitude. This observation supports earlier conjecture of emergent inhomogeneity of the conductive
regions near $n_c$ \cite{morgun_PRB_2016} deduced from the analysis of  magnetoconductivity in weak parallel fields.
Thus, we associate the observed oscillations with Shubnikov-de Haas (SdH) effect within certain regions of the 2D space, in which the momentum relaxation times $\tau_p$ are much longer than that calculated from the global resistivity under assumption of the uniform current flow.

\begin{figure}[ht]
	\includegraphics[width=210pt]{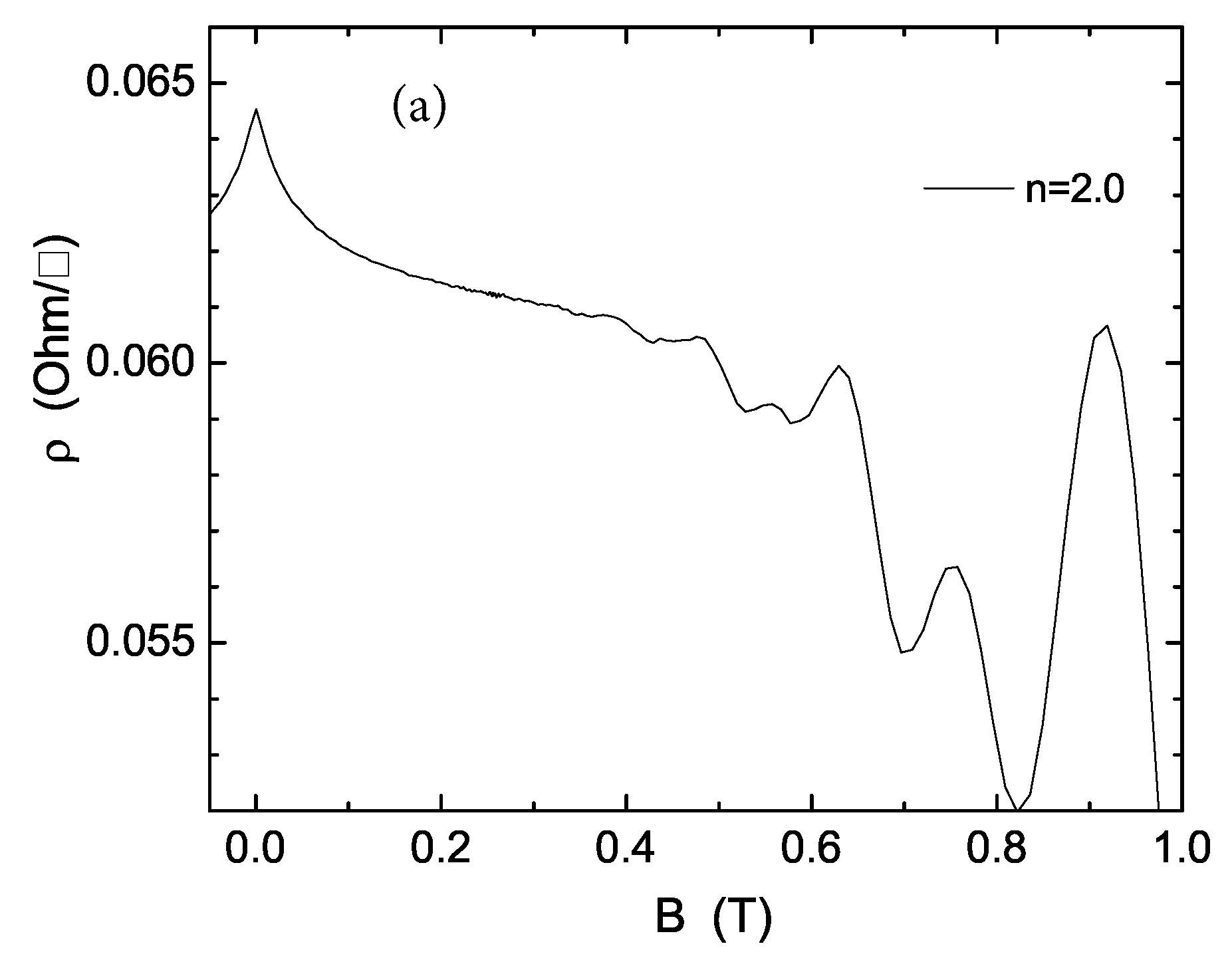}
		\includegraphics[width=210pt]{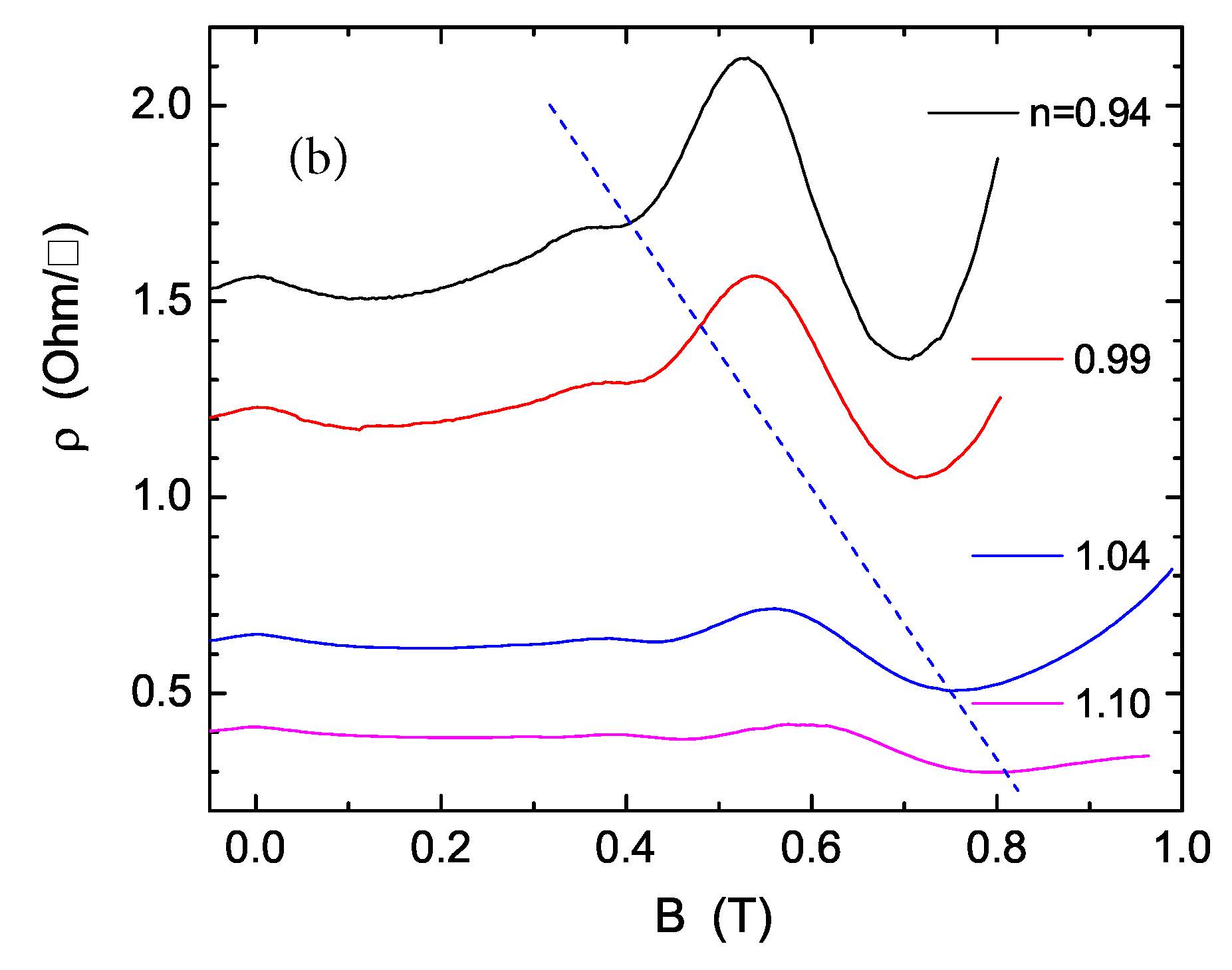}
	\caption{Examples  of quantum oscillations of the resistivity for (a) $n=2\times 10^{11}$cm$^{-2}$,
	and (b) 0.94, 	1.00, and $1.04\times 10^{11}/$cm$^2$.
 Dashed lines in panel (b) depict the upper boundary of analyzed magnetic fields. $T= 0.1$K.
 Adapted from \cite{pudalov-gersh_JETPL_2020}.}
	\label{fig:SdH-low_n}
\end{figure}

In Ref.\cite{teneh_PRL_2012} it was shown that the correlated 2D  electron system can be inhomogeneous even {\em  at high  electron densities}: it contains inclusions of collective  localized (insulating) states (called ``spin droplets'') in a  conductive Fermi liquid. From the low density  oscillatory transport measurements \cite{pudalov-gersh_JETPL_2020} the latter  picture (we call it ``bi-colored'') is supplemented with
 the data showing that the system is, in fact, ``three-colored''.
 The conductive Fermi-liquid phase is not spatially  homogeneous. Instead, it forms a pattern of regions
 with a large momentum relaxation time $\tau_p$. These  highly conductive regions are connected with each
 other through poorly conductive regions of Fermi liquid  with lower $\tau_p$-values.
 
\subsubsection{Phase separation effects in  zero field transport}

\begin{figure}[ht]
\includegraphics[width=240pt]{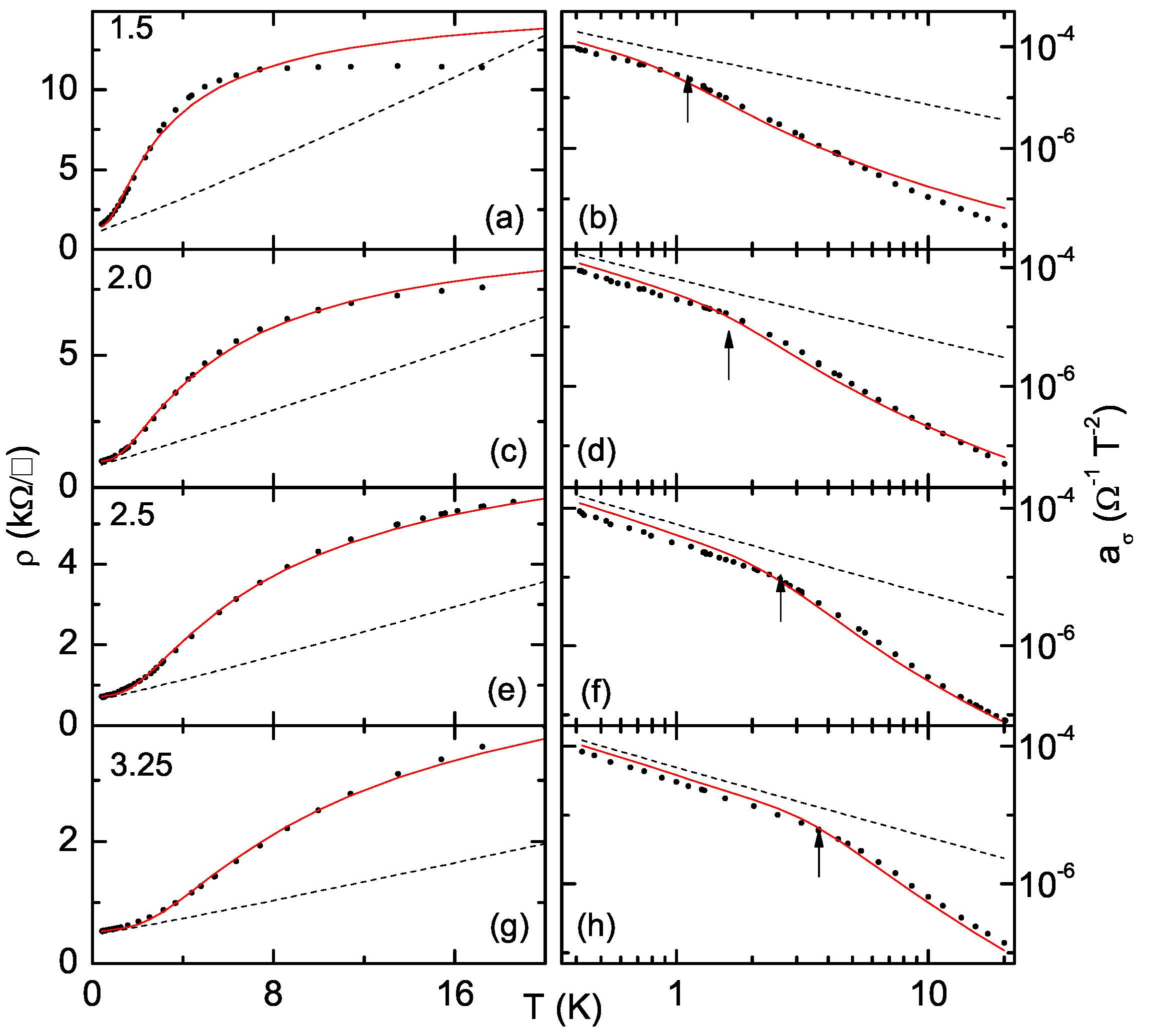}
	\caption{Fitting $\rho(T,B = 0)$ dependencies (left) and $a_\sigma (T)$ (right)
		with the same set of the fitting parameters. Carrier
		densities (from top to bottom) are $n = 1.5, 2.0, 2.5$, and $3.25\times
		10^{11}$cm$^{-2}$.  Vertical arrows
		point at the kink positions. Adapted from  \cite{morgun_PRB_2016}.	}
	\label{fig:dualFit}
\end{figure}

Below we analyze the   $\rho(T)$ and $\sigma(T)$ dependencies at zero field. The variations of
these quantities in the relevant temperature range  (see Fig.~\ref{fig:dualFit}) for high mobility Si-MOS samples  are  large (up to a factor of  10), making the IC theory inapplicable.

Each $\rho(T)$ curve has two remarkable  points:  $\rho(T)$ maximum, $T_{\rm max}$, and
inflection,  $T_{\rm infl}$ \cite{knyazev_JETPL_2006, knyazev_PRL_2008}.  Whereas $T_{\rm max}$ is an order of the renormalized Fermi
energy, the  inflection point happens at much lower temperatures, in the degenerate regime.
Importantly, the inflection temperature 
appears to be close to the kink temperature (see Figs.~\ref{fig:phase diagram}, \ref{fig:dualFit}).
Besides that, $T^*(n)$ is much higher than   the ``incoherence'' temperature at which the phase coherence is lost
(defined as $\tau_\varphi(T) =\tau$  \cite{brunthaler_PRL_2001}).
This confirms that the kink, inflection and $\partial \chi/\partial n$ sign change are irrelevant to the single-particle
interference effects \cite{brunthaler_PRL_2001, pudalov-lnT_JETPL_1998, pud_PhysE_1998, klimov_PRB_2008}.

One can see from  Fig.~\ref{fig:dualFit}  that the $\rho(T)$ temperature dependence is  monotonic up to 
$T=T_F$, and follows one and the same additive resistivity functional form  over a 
wide density range:
\begin{eqnarray}
\rho(T) &=&\rho_0 +\rho_1\exp(-\Delta(n)/T),\nonumber \\
	\Delta (n)&=& \alpha (n-n_c(B)),
	\label{eq:rho-exp}
\end{eqnarray}
where  $\rho_1(n,B)$ is  a slowly decaying function of $n$, and $\rho_0(n,T)$ includes Drude resistivity and quantum
corrections, both from the single-particle interference and interaction.
Although the above empirical  resistivity form has been suggested in Ref.~\cite{pudalov-SO_JETPL_1997} on a different footing, it fits well the $\rho(T)$ dependence for a number of  material systems
\cite{pudalov-SO_JETPL_1997, hanein_1998, papadakis_1998, gao_2002,
	brunthaler_SOI_2010, zhang-SrTiO3_PRB_2014, raghavan-SmTiO3_APL_2015, pouya-SrTiO3_PRB_2012}.

This empirical  additive $\rho(T)$ form  satisfies general requirements for the transport behavior in the vicinity of
a critical point \cite{amp_2001, knyazev_PRL_2008}.
This form implies  two channel scattering and therefore  agrees with the two-phase state of the 
low density 2D electronic system (cf. Matthiessen's rule). 

As noted above,  $\rho(T)$ (and $\sigma(T)$) variations of the experimental data
(Fig.~\ref{fig:dualFit}) are so large, that the first order in $T$ corrections, of cause, cannot describe them.
The simplest functional dependence, Eq.~(\ref{eq:rho-exp}),   correctly describes  the inflection in $\rho(T)$ and
the linear density dependence of the inflection temperature   \cite{pudalov-SO_JETPL_1997, pudalov_disorder_2001}.
Obviously, in this model $T_{\rm infl}= \Delta/2$. To take magnetic field into account, and following the results of
Ref.~\cite{pudalov_disorder_2001} 
we include to $(\Delta/T)$ all  even in $B$ and the lowest order in $B/T$  terms, as follows:
\begin{equation}
	\Delta(T,B,n)/T =\Delta_0(n)/T - \beta(n) B^2/T  - \xi(n) B^2/T^2,
	\label{eq:nc(B)}
\end{equation}
with $\Delta_0=\alpha[n-n_c(0)]$.

Equations~(\ref{eq:rho-exp}) and (\ref{eq:nc(B)}) link the magnetoconductance  with the zero-field $\rho(T)$
temperature  dependence. Combining  equations (\ref{eq:rho-exp}) and (\ref{eq:nc(B)}), we obtain the $\rho(T,B)$ dependence as follows:
\begin{eqnarray}
	&\rho(B,T) = \left[\sigma_D -
	\delta\sigma \cdot \exp \left(- T/T_B \right)\right]^{-1} \nonumber \\
	&+ \rho_1 \exp \left( - \alpha \frac{n-n_c(0)}{T} - \beta \frac{B^2}{T} - \xi \frac{B^2}{T^2}\right)
	\label{eq:r(B&n)}
\end{eqnarray}
The term in the square brackets includes the Drude conductivity and interaction quantum corrections
\cite{ZNA-R(T)_PRB_2001, ZNA_BPar_2001}, which are smoothly cut-off 
above $T=T_B\approx \Delta/2$. $\delta\sigma(T)$
was calculated in Ref. 
 \cite{morgun_PRB_2016} using experimentally determined Fermi-liquid coupling constants $F_0^\sigma (n)$ 
\cite{gm, klimov_PRB_2008}, and $\sigma_D$ found by a conventional  procedure
\cite{pudalov_R(T)_PRL_2003}. 

From Eq.~(\ref{eq:r(B&n)}),
the prefactor  $a_\sigma = - (1/2)\partial^2\sigma/\partial B^2 $ is calculated straightforward and
in Fig.~\ref{fig:dualFit} is compared with experimental data.
In the $\rho(T)$ fitting [Figs.~\ref{fig:dualFit}\,(a,c,e,g)],  basically, there is only one adjustable parameter,
$\rho_1(n)$, for each density. Indeed, 
$n_c(0)$ is determined from the conventional scaling analysis at $B=0$ \cite{knyazev_PRL_2008}, and the slope
$\alpha = 2 \partial T_{\rm infl}(n)/\partial n$
may be determined  from Fig.~\ref{fig:phase diagram}.

One can see that both $\rho(T)$ and $a_\sigma(T)$ are well fitted; the model captures correctly the major data
features, the steep $\rho(T)$  rise (including the inflection), and the kink in  $a_\sigma(T)$ dependence. Within this  model, the kink 
signifies a transition from the low-temperature magnetoconductance regime
(where  the interactions driven linear $\sigma(T)$ temperature dependence 
dominates and the exponential term may be neglected) to the high temperature regime governed
by the steep exponential $\rho(T)$ rise. 

We emphasize that both regimes are not related to the diffusive regime of interactions. This conclusion
casts doubt on early attempts to use the two-parameter scaling for describing the magnetoresistance  $\sigma(B)$ and the temperature dependence  $\rho(T)$ within the renormalization group approach.

Thus, within the framework of a phenomenological two-phase model with two scattering channels, it is possible to explain all the observed features in transport and magnetotransport in a parallel field. It is important that their characteristic temperatures are close to the crossover temperature $T_{dM/dn}(n)$, where the spin magnetization per electron changes the sign \cite{teneh_PRL_2012} (see the insert in Fig.~\ref{fig:phase diagram}).
Physically, this means that at temperatures below $T_{dM/dn}(n)$, collective droplets with a large spin (minority phase) ``melt'' with increasing density. In other words, the electrons added to the Fermi liquid improve screening thereby promoting the disappearance of spin droplets. At temperatures above $T_{dM/dn}(n)$, on the contrary, the number of spin droplets increases with increasing density; in this case, the electrons added to the 2D system prefer to combine and form new spin droplets. In Ref.~\cite{morgun_PRB_2016}, it was concluded that $T^*$ may be related to the averaged energy spectrum of the SD phase.

 \subsection{Phase separation effect in spin susceptibility}          

Using a vector  magnetic field technique 
with two independent superconducting coils, in Ref.~\cite{pudalov-nonlinear_PRB_2021}  SdH oscillations were precisely measured and analyzed in various in-plane fields. Earlier \cite{gm, pudalov_PRB_2018, pudalov-spinless}, the oscillatory component $\delta \rho_{xx}$ was shown   to be  well fitted with conventional Lifshits-Kosevich formula \cite{SdH, isihara_1986, pudalov-spinless,  pudalov_PRB_2018}; 
this enables accurate extraction of the spin susceptibility $\chi^*$ and density of mobile carriers $n_{\rm SdH}$ from the beating of oscillations. In particular, $\chi^*$ values have been determined  with an accuracy of   $\sim (1 - 2)\%$ as a function of the in-plane field.
         
\begin{figure}[ht]
\includegraphics[width=200pt]{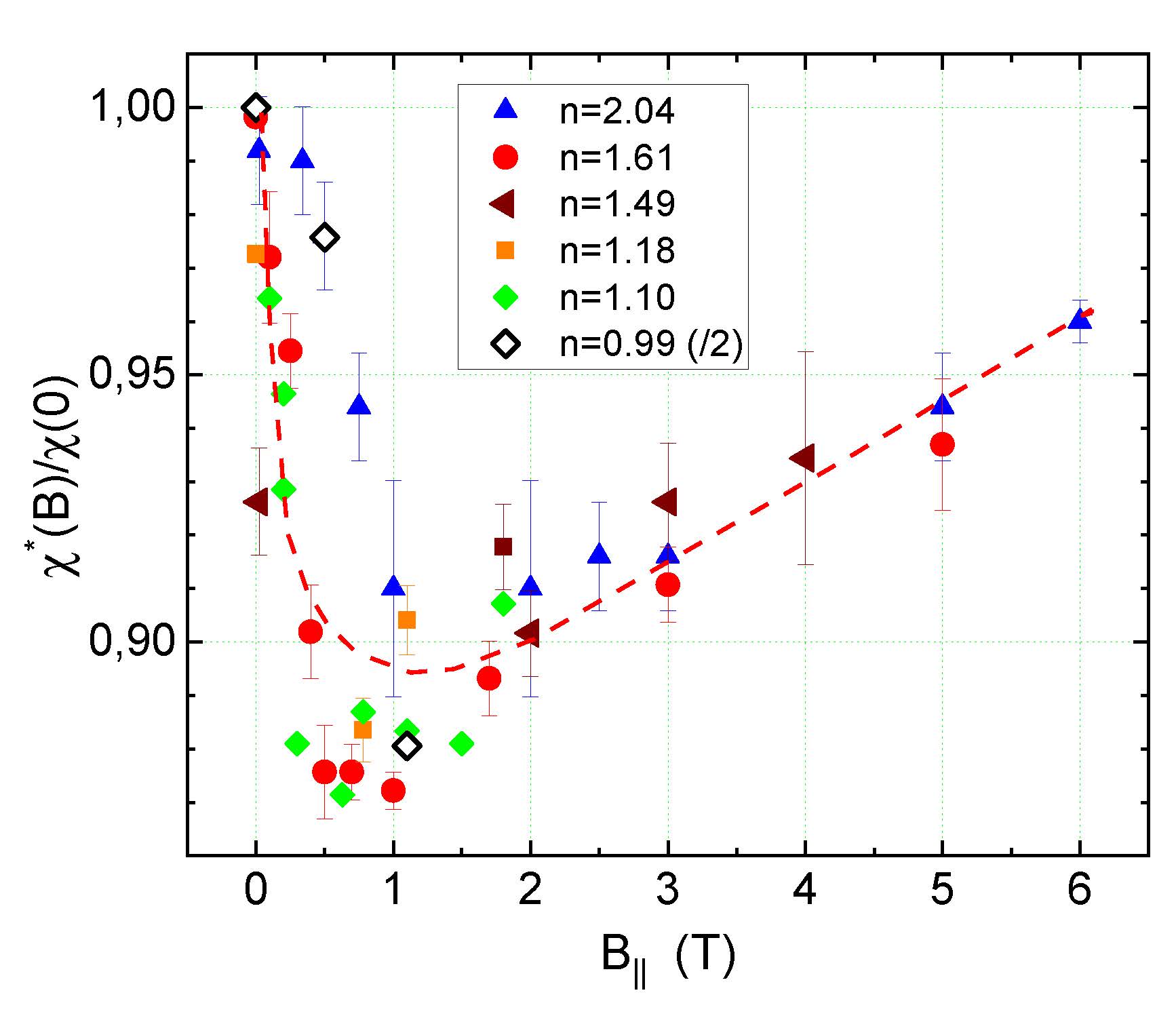}
	\caption{
		Summary of $\chi^*(B_\parallel)/\chi^*(0)$ data versus $B_\parallel$  for two samples and for several densities.
		For the lowest density $n=0.99$, the $\chi^*(B)/\chi^*(0)$ variations are scaled down by 2 times.
		The density is indicated in units of $10^{11}$cm$^{-2}$,  $T=0.1$\,K. Adapted from Ref.~\cite{pudalov-nonlinear_PRB_2021}}
	\label{Fig1-nonlinear}
\end{figure}

Figure \ref{Fig1-nonlinear}  shows  the main result of \cite{pudalov-nonlinear_PRB_2021} - a sharp nonmonotonic dependence of $\chi^*$ on the in-plane field. The characteristic $\delta\chi^*(B)/\chi^*(0)$ variations  are in the range from   $\sim 25\%$ at low densities to  $\sim 6\% $ at high densities $10\times 10^{11}$cm$^{-2}$.
The data reported in \cite{pudalov-nonlinear_PRB_2021}   coincide in the $B_\parallel \rightarrow 0$ limit with the $\chi^*(B=0)$ values reported Refs.~\cite{gm, klimov_PRB_2008}.
The characteristic field of the  $\chi^*(B)$-minimum, $B_\parallel \sim 1$\,T for $n=(1.1-2)\times 10^{11}$cm$^{-2}$,  is much weaker than the field of complete spin polarization of the 2D system 
$B_p\sim 20$T. 
Evidently, in a homogeneous single-phase Fermi liquid the only characteristic field is $B_p$.

\begin{figure}
\includegraphics[width=200pt]{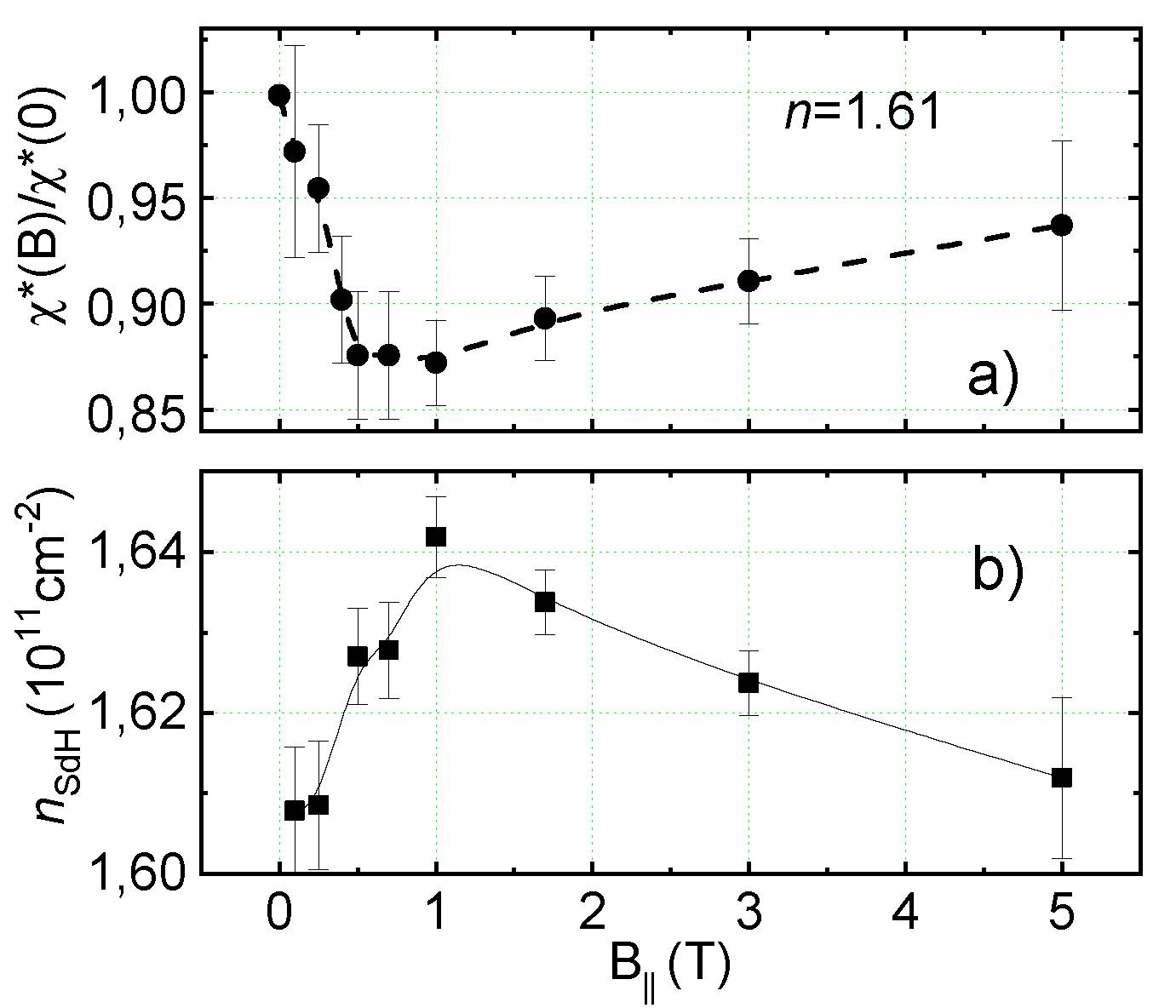}
	\includegraphics[width=200pt]{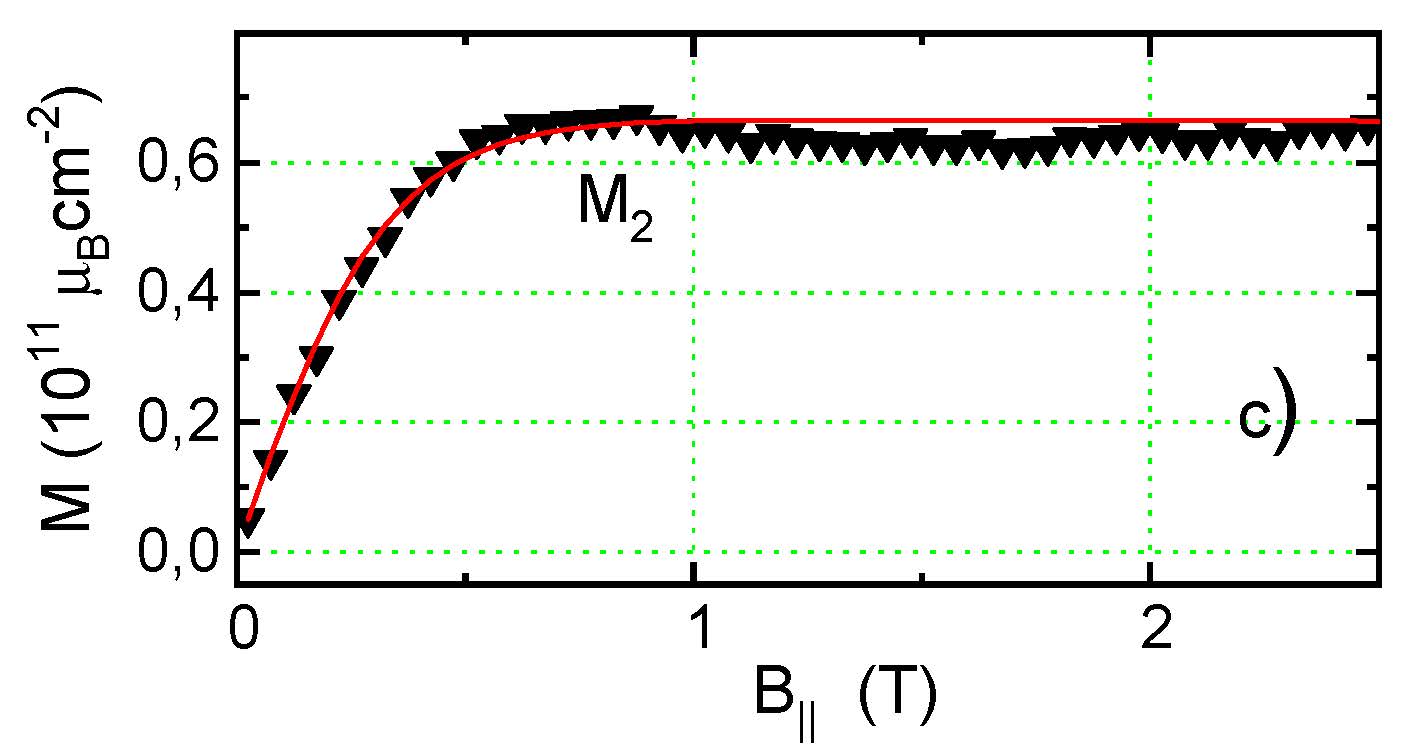}
	\caption{Correlation between  the in-plane field dependence of (a) $\chi^*(B)/\chi^*(0)$, and 
		(b) density $n_{\rm SdH}(B)$.
		Red curve  shows $\tanh(\mu_B B/k_B T)$-fitting of the experimental $M(B)$ data.
	The zero-field densities are  $n_0=1.61\times 10^{11}$cm$^{-2}$ for (a) and (b), and $1.4\times 10^{11}$cm$^{-2}$ for (c).
	 Temperature $T=0.1$\,K. Adapted from Ref.~\cite{pudalov-nonlinear_PRB_2021}.
	}
	\label{Fig2ab-nonlinear}
\end{figure}

The spin susceptibility variations $\delta\chi^*(B)$  measured from  SdH oscillations are relevant to
the mobile carriers. The $\delta\chi^*(B)$ data also appears to correlate (i)  with  variation of the mobile carrier density $\delta n_{\rm SdH}$ (see Fig.~\ref{Fig2ab-nonlinear}), and (ii) with thermodynamic magnetization $\partial M/\partial n$ of the collective localized states $M(B)$ (see Fig.~\ref{teneh_PRL_2012}) \cite{teneh_PRL_2012}.
This correlation prompts that the observed changes in the properties of extended states are caused by the  changes in magnetization of the localized states  and by the subsequent redistribution of carriers between the two subsystems.
The range of accessible densities where $\delta n_{\rm SdH}$ could be measured is limited from low densities side,   on the verge of the transition to fully localized state. Here the $n_{\rm SdH}(B)$ variation cannot be measured precisely and variations of $\chi^*(B)$ cannot be traced to higher field, because application of an in-plane field quickly  causes localization of the 2D system  \cite{simonian-MR_PRL_1997, pud-MR_JETPL_1997, pud_PhysicaB_1998, krav-tilted_PRB_1998}

\begin{figure}
	\hspace{0.25in}\includegraphics[width=160pt]{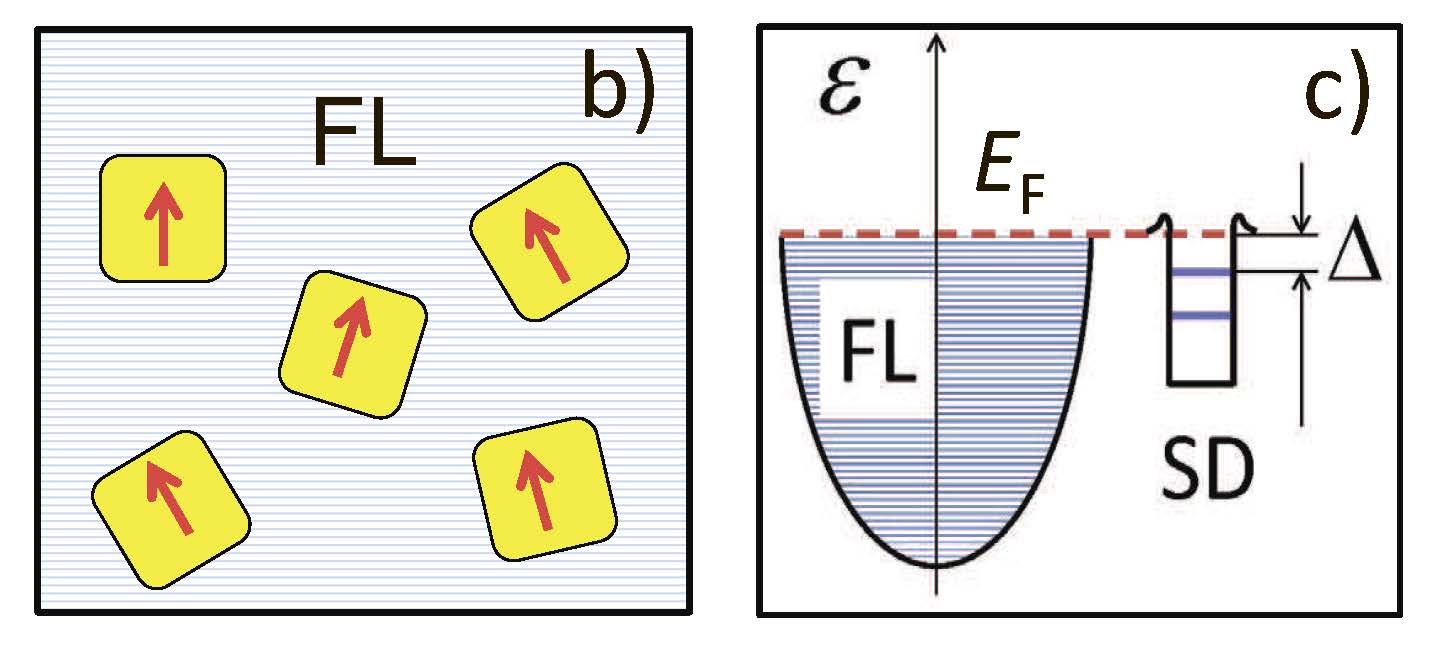}
	\caption{(a) Schematic spatial arrangement of the two-phase state and (b)  the energy band diagram of the two-phase system.
	}
	\label{fig6-nonlinear}
\end{figure}

From the density redistribution at ultralow temperature it follows that energy of the localized states is located in the close vicinity of the Fermi energy, in order to 
allow  for the  carrier exchange  at ultralow temperatures between two electronic subsystems.
No temperature dependence of $\delta n_{\rm SdH}$ was observed within the range $0.1-0.5$\,K, hence,  the carrier redistribution occurs elastically, via tunneling.  The  energy diagram describing schematically the two-phase state is
shown in Fig.~\ref{fig6-nonlinear}.  Note, that this picture is essentially different from the conventional model of the disorder-localized single-particle  states  in the tail of the conduction band \cite{ando_review, gold_JPCM_2002, 
vitkalov_PRB_2002}.

It is worthnoting  that the Fermi-liquid density  deduced from SdH oscillations in the phase-separated system is determined by the local density in the Fermi-liquid ``lakes'' (where the carriers possess the highest relaxation time), rather than by the total or by average  density; this picture holds until the delocalized states (Fermi-liquid lakes) percolate. 
The carrier redistribution between two phases in the 2D system  is not easy to determine by other techniques. For example, the capacitance measurements taken at frequencies 
$10^1 - 10^5$\,Hz (1\,nF, 10\,kOhm/$\Box$)
probe the total charge density that includes both SD and mobile states.  In order to separate the SD and FL states, the capacitance measurements should be done at frequency of ~$10^{10}-10^{12}$Hz, inaccessible for the gated structure.
Hall measurements cannot also shed a light on the density distribution of the delocalized and  SD states, because Hall 
voltage becomes irrelevant to the carrier density in the vicinity of the localization transition   
\cite{pudalov_JETPL_1993}.

Measurements \cite{pudalov-nonlinear_PRB_2021} have been performed with a gated Si-MOS structure at a fixed gate voltage $V_g$, whereas 
$B_\parallel$ and $T$  varied. Under this condition the total charge is conserved.
Therefore, a change in the mobile carrier density ($\delta n_{\rm SdH}$) in the FL-regions can only occur via carriers transfer to the localized regions (SD) and back \cite{teneh_PRL_2012, tupikov_NatCom_2015}.

\begin{figure}
\includegraphics[width=200pt]{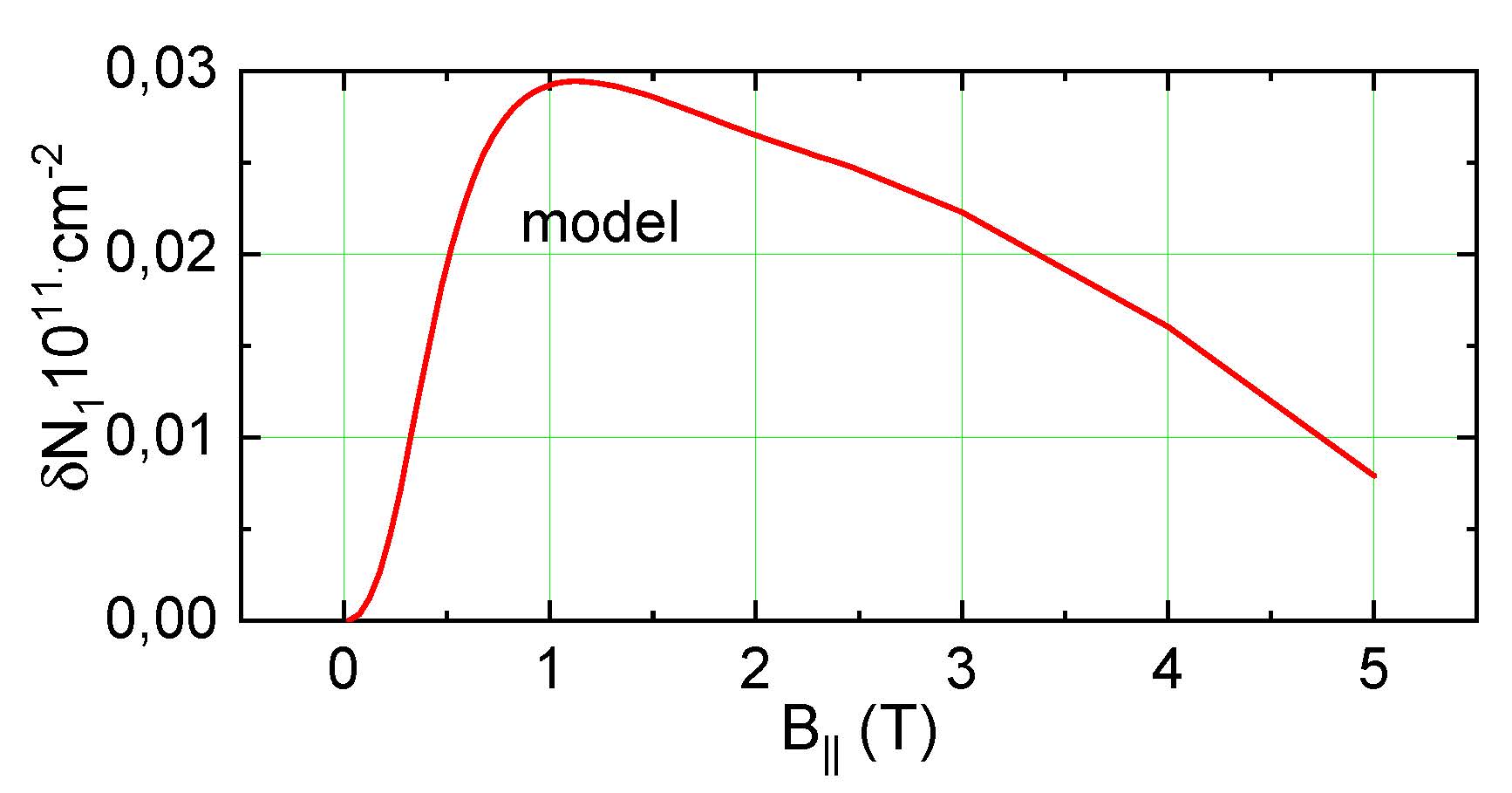}
	\caption{Model curve $\delta N_1(B_\parallel)$ calculated from experimental data as described in the text.
		Adapted from Ref.~\cite{pudalov-nonlinear_PRB_2021}.
	}
	\label{fig7-nonlinear}
\end{figure}

To describe the data, 
a simple  thermodynamic model with two  phases coexisting in equilibrium  has been applied in Ref.~\cite{pudalov-nonlinear_PRB_2021}.
The model  was found is capable to explain the results qualitatively, and even quantitatively, with some parameters determined in experiments.  In particular, the $n_{\rm FL}(B)$ dependence    calculated  
within this model  for a representative density $1.4\times 10^{11}$cm$^{-2}$
is shown in Fig.~\ref{fig7-nonlinear}. 
It is rather similar to the direct experimental data of Fig.~\ref{Fig2ab-nonlinear}b; the similarity supports
the validity of the two-phase thermodynamic approach.
          
Summarizing the content of this section, we conclude that the results of \cite{morgun_PRB_2016, pudalov-nonlinear_PRB_2021} give reason to believe that the phase separation in a correlated 2D electron system exists not only near the transition to the insulator state (as was revealed in local compressibility measurements \cite{ilani_PRL_2000}), but also in a wide range of densities, even deep in the ``metallic regime" of high conductivity $\sigma = (3 - 80)\times(e^2/h)$ \cite{gmax}.

\section{Conclusions}
A two-dimensional electron system in silicon structures for the last 50 years has served as a research platform, 
where many new exciting effects have been discovered, including the integer quantum Hall effect,
negative electron compressibility, strong renormalization of electron effective mass and spin susceptibility, etc.
This 2DE system is strongly correlated in a wide range of densities,
where the energy of interparticle interactions is much greater than the kinetic Fermi energy.

Local compressibility measurements \cite{ilani_PRL_2000} evidenced the emergence of an inhomogeneous state on a microscopic scale in 2D system with a decrease in the concentration of carriers near the transition to the insulator state. For a long time, this result was not appreciated when considering a macroscopic system with high conductivity as, on average, a homogeneous Fermi liquid. 
Within such approach, the averaged values of the  Fermi-liquid parameters were experimentally determined  and the averaged properties in charge transport were quantitatively described. However, later thermodynamic measurements \cite{teneh_PRL_2012} revealed signatures of the coexistence in thermodynamic equilibrium, in a wide range of densities, of the majority Fermi liquid and the minority phase of collective localized states with large spin.

Subsequent precision measurements of SdH oscillations in the presence of an in-plane  field revealed a sharp change in the spin susceptibility $\chi^*(B_\parallel)$ and a simultaneous change in the concentration of mobile carriers $\delta n_{\rm SdH}(B)$ 
in  correlated 2D electron system. 
 The two effects correlate well with each other and with the thermodynamic magnetization of the localized SD states. 
 It is found that the origin of these variations is the magnetization of collective localized states (``spin droplets'') and, as a result, the redistribution of carriers between the two phases. Independent measurements of spin magnetization and magnetoresistance in a weak in-plane field, as well as the temperature dependence of resistance, revealed the existence of a new energy scale $T^*(n)\ll T_F$, which marks a crossover between the regime of predominant proliferation of the SD states and the regime of their disappearance.
 The results of the considered experiments were described within the framework of a phenomenological two-phase model. These results  and their susccessfull description with two-phase model provide the solid evidence for the  phase separation in the interacting 2D electron system even at relatively high carrier densities, deeply in the ``metallic'' regime of high conductivity $\sigma = (3 - 80)\times(e^2/h)$ \cite{gmax}. The latter regime was commonly considered as a pure Fermi liquid.         

In quasi-one-dimensional systems, the main driving force of the phase separation is associated with the nesting of the Fermi surface,
which leads to the appearance of a spin or charge density wave coexisting with a paramagnetic or superconducting metallic phase in the vicinity of the phase transition \cite{lebed-book, kornilov_PRB_2004, gerasimenko_PRB_2014}.
In 2D systems, instability can also occur in the charge or spin exchange channel. An interesting and still debatable issue is the microscopic mechanism behind the electronic phase separation that is  experimentally observed in  correlated low dimensional electron systems.

  Several scenarios were theoretically considered, where in  the majority Fermi liquid the minority phases  such as Wigner solid ``droplets'', or spin polarized ``droplets'' emerge due to the local Wigner crystallization \cite{spivak_2003, spivak_2004}, local Stoner instability \cite{eisenberg_PRB_1999, kurland_PRB_2000, narozny_PRB_2000, benenti_PRL_2001, slogett_PRB_2005, stadnik-sushkov_2013, repin-burmistrov_PRB_2021}, or, alternatively,  the topology of the Fermi surface changes \cite{khodel_JETPL_1990, volovik_JETPL_1991, khodel_PRB_2008, dolgopolov_UFN_2019, dolgopolov_JETPL_2022}. 

The experimental results presented in this review  evidence for the existence of spin-polarized droplets as the minority phase in the majority Fermi liquid sea. It is possible, however, that with a stronger interaction or a weaker disorder, instability in the charge channel may also manifest itself. 

Attempts to ignore the tendency to spin/charge instability or instability of the Fermi surface, considering only the semiclassical effects of disorder and screening \cite{das_PRB_2013}, although they are able to describe some experimental results (such as negative compressibility, transport in the zero field), but give an overly simplified picture of the phenomenon of phase separation and miss the structure of the heterophase state.
  
The microscopic mechanism responsible for the phase separation, for the redistribution of carriers between two phases, as well as the energy structure of the minority phase remain interesting and still open issues.
                
\section{Acknowledgements}
The author is grateful to  B. Altshuler, G. Bauer, G. Brunthaler, I.S. Burmistrov,   M. D'Iorio, J. Campbell, V.S. Edel'man, M.E. Gershenson,  N. Klimov, H. Kojima, S. V. Kravchenko, 
A. Yu. Kuntsevich,  D. L. Maslov, L. A. Morgun, O. Prus, M. Reznikov, D. Rinberg, S. G. Semenchnisky,  and N. Teneh 
for fruitful collaboration in developing experimental methods, performing  measurements,  discussing the results, and writing the original papers. Financial support from the State assignment of the research at P.N. Lebedev Physical Institite (Grant \# 0019-2019-0006) 
 and from Russian Foundation for Basic research  (\#18-02-01013) is acknowledged.

\end{document}